\documentclass[a4paper, useAMS,usenatbib,usedcolumn]{mn2e}	%Two columns
\usepackage[figuresright]{rotating}
\usepackage{lscape} 
\pdfoutput=1
\usepackage{graphics}

\usepackage{bigdelim}
\usepackage{bigstrut}

\usepackage{setspace}

\usepackage{natbib}
\usepackage{mathtools}
\usepackage{amsmath}
\usepackage{amssymb}
\usepackage{pifont}
\usepackage[subnum]{cases}

\usepackage{soul}
\newcommand{\sauron}{{{SAURON}}}
\newcommand{\atlas}{{{ATLAS$^{\rm{3D}}$}}}

\newcommand{\placetabA}	
{	%%Galaxy sample
	\begin{table*}
	\begin{center}
	\begin{tabular}{cccccccc}
		\hline
		
		Galaxy 		& $\rm{R_{e}}$	& Type			& Dist.		& $\rm{\sigma_{\rm{kpc}}}$ 	& $M_{B}$	& $log(\rm{M_{\star}/M_{\odot}})$ 	& $(g-i)$	\\
					&	(arcsec)	& 				& (Mpc)		& ($\rm{km~s^{-1}}$)	& ($\rm{mag}$)	& ($\rm{dex}$)			& ($\rm{mag}$)	\\		
		(1)			&	(2)			& (3)			& (4)		& (5)					& (6)			& (7)					&	(8) 	\\		
		
		\hline
		NGC~1023	& 48			& S0			& 11.1		& 183					& -20.15		& 10.92					& 0.96 		\\ 	
		NGC~1400	& 28			& E1/S0			& 26.8		& 236					& -20.36		& 11.03					& 0.95 		\\	
		NGC~1407	& 63			& E0			& 26.8		& 252					& -21.61		& 11.50					& 0.98 		\\  
		NGC~2768	& 63			& E6/S0			& 21.8		& 206					& -21.02		& 11.20					& 0.95 		\\
		NGC~3115	& 35			& S0			& 9.4		& 248					& -20.10		& 10.91					& 0.93 		\\	
		NGC~3377	& 36			& E5--6			& 10.9		& 135					& -19.02		& 10.42					& 0.93 		\\	
		NGC~4278	& 32			& E1--2			& 15.6		& 228					& -19.99		& 10.83					& 0.86 		\\	
		NGC~4365	& 53			& E3			& 23.1		& 253					& -21.30		& 11.40					& 0.91 		\\	
		NGC~4473	& 27			& E5			& 15.2		& 189					& -19.82		& 10.82					& 0.99 		\\	
		NGC~4494	& 49			& E1--2			& 16.6		& 157					& -20.45		& 10.96					& 0.99 		\\	
		NGC~4649	& 66			& E2/S0			& 16.5		& 308					& -21.32		& 11.46					& 0.99 		\\	
		NGC~5846	& 59			& E0--1/S0		& 24.2		& 231					& -21.02		& 11.31					& 0.96 		\\
		\hline
	\end{tabular}
	\end{center}
		\caption{Galaxy parameters. 
		The columns present: 
		(1) Galaxy name. 
		(2) Effective radius. 
		(3) Morphological type, (4) Distance and (5) central velocity dispersion 
		from \citet{Brodie14}.  
		(6) Total B-band magnitude from \citet{deVaucouleurs91}, corrected 
		for galaxy extinction. 
		(7) Total logarithmic stellar mass obtained from the $K$-band extinction corrected absolute 
		magnitude from the 2MASS source catalogue \citep{Jarrett00}, obtained following the 
		same approach as in \citet{Pastorello14}, assuming the distances in \citet{Brodie14}. 
		(8) Colour marking the split between the blue and red GC subpopulations 
		from \citet{Usher12}, except for NGC~1023, NGC~4473 and NGC~4649 
		(calculated in this work). 
		}			
	\label{tab:A}
	\end{table*}
}

\newcommand{\placetabD}	
{	%%Galaxy sample
	\begin{table}
	\begin{center}
	\begin{tabular}{ccc}
		\hline

									& Low-mass $\nabla[Z/$H]	& High-mass $\nabla[Z/$H] \\	
									& ($\rm{dex/dex}$)	 						&  ($\rm{dex/dex}$)		\\	
		\hline
	rGC								& $-0.26\pm0.08$								& $-0.07\pm0.08$ \\	
	bGC								& $-0.45\pm0.13$								& $-0.08\pm0.09$ \\	
		\hline
	\end{tabular}
	\end{center}
		\caption{GC metallicity gradients in the two galaxy mass bins with 
		separation at $M_{\rm{\star}}=10^{11}~\rm{M_{\odot}}$. 
		The gradients are measured in the whole available radial range 
		(i.e. $0.6\lesssim R\lesssim14~\rm{R_{e}}$) 
		}	
	\label{tab:D}
	\end{table}
}

\newcommand{\placetabC}	
{	%%Galaxy sample
	\begin{table}
	\begin{center}
	\begin{tabular}{ccc}
		\hline
		
		Galaxy 		& $m$	& $q$	\\%	& $m_{\rm{bGC}}$	& $q_{\rm{bGC}}$	\\
					&		& (dex) 			\\%& 					& (dex)				\\	
		\hline
		NGC~1407	& $-0.01\pm0.02$	& $-0.31\pm0.08$	\\%& $-0.03\pm0.01$	& $-0.97\pm0.09$	\\ 	
		NGC~3115	& $0.00\pm0.01$		& $-0.18\pm0.10$	\\%& $-0.04\pm0.02$	& $-1.07\pm0.12$	\\ 
		NGC~4278	& $-0.02\pm0.01$	& $-0.40\pm0.08$	\\%& $0.01\pm0.01$		& $-1.45\pm0.13$	\\ 
		NGC~4365	& $-0.02\pm0.02$	& $-0.37\pm0.07$	\\%& $0.00\pm0.02$		& $-1.20\pm0.12$	\\ 
		NGC~4649	& $0.03\pm0.04$		& $-0.38\pm0.12$	\\%& $0.02\pm0.03$		& $-0.96\pm0.13$	\\ 
		\hline
	\end{tabular}
	\end{center}
		\caption{Best linear fit parameters for the rGC metallicity profiles of 
				the five galaxies with the highest number of rGCs. 
				The linear fit is in the form: $[Z\rm{/H}]=m \log(R\rm{/R_{e}}) + q$
		}			
	\label{tab:C}
	\end{table}
}

\newcommand{\placetabAppData}	
{	
	\begin{table*}
	\begin{center}
	\begin{tabular}{lcccccc}
		\hline
		
		Name		& R.A.		& Dec. 	& ($g-i$) 	& $i$ 	& CaT 			& $[Z/$H]	\\
					& [deg]		& [deg]	& [mag]		& [mag]	& $\rm{\AA}$	& [dex]		\\
		(1)			& (2) 		& (3)	& (4)		& (5)	& (6)			& (7)		\\
		\hline

		NGC1023\_GC1	& $40.1543$	& $39.0783$	&  $1.08\pm0.03$	& $21.98\pm0.01$	& $7.64^{+0.82}_{-1.01}$ & $-0.23^{+0.38}_{-0.46}$ \\
		NGC1023\_GC2	& $40.1542$   & $39.0645$	& $1.07\pm0.01$	& $19.71\pm0.01$	& $6.49^{+0.22}_{-0.18}$ & $-0.76^{+0.10}_{-0.08}$ \\
		NGC1023\_GC3 	& $40.1460$	& $39.0370$	& $1.12\pm0.06$	& $22.72\pm0.04$	& $8.24^{+0.68}_{-1.78}$ & $0.05^{+0.31}_{-0.82}$ \\

		$\dotso$	& $\dotso$	& $\dotso$	& $\dotso$	& $\dotso$	& $\dotso$	& $\dotso$	\\
		\hline
	\end{tabular}
	\end{center}
		\caption{GC CaT and metallicity measurements summary. 
		The full version of this table is provided in a machine readable form in 
		the online Supporting Information. 
		The columns present: 
		(1) Globular cluster IDs.  
		(2) and (3) Right ascension, declination in the J2000.0 epoch. 
		(4) ($g-i$) colour. 
		(5) $i$ magnitude. 
		(6) CaT index measurement. 
		(7) Total metallicity. 
		}	
	\label{tab:AppData}
	\end{table*}
}

\newcommand{\placefigAfirst}{
	\begin{figure*}
    	\begin{center}
			\includegraphics[width=2\columnwidth]{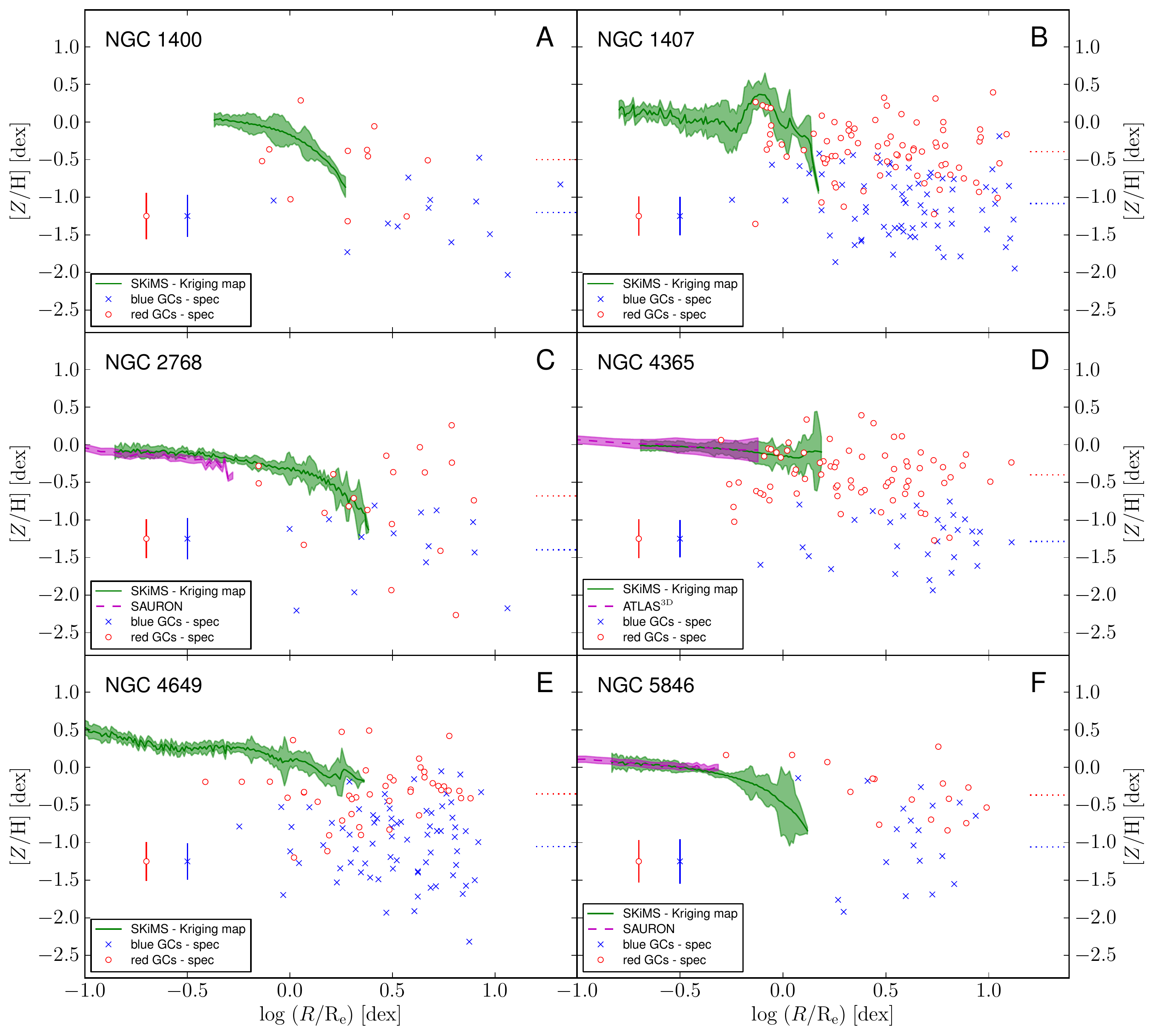}
		\end{center}
	    \caption[]{
	    		Metallicity profiles for the high-mass galaxies in our sample. 
	    		The panels (A-F) show metallicity versus elliptical
	    		galactocentric radius scaled by $\rm{R_{e}}$. 
	    		The SKiMS stellar metallicity profiles are presented as green 
	    		solid lines. 
	    		When available, the \sauron\ stellar metallicity radial profiles 
	    		are presented as a magenta dashed line. 
	    		For NGC~4365 (panel D), the magenta dashed line shows instead 
	    		the \atlas\ stellar metallicity profile. 
	    		Blue crosses and red circles show, respectively, blue and red 
	    		GC metallicities. 
	    		The blue crosses and red circle points with errorbars show the 
	    		average uncertainties on the metallicities of blue and red GCs, 
	    		respectively. 
	    		The red and the blue dotted lines show the average 
	    		metallicities of the rGC and bGC subpopulations, respectively. 
	    	    }
    \label{fig:A1}
  \end{figure*}
}

\newcommand{\placefigAsecond}{
	\begin{figure*}
    	\begin{center}
			\includegraphics[width=2\columnwidth]{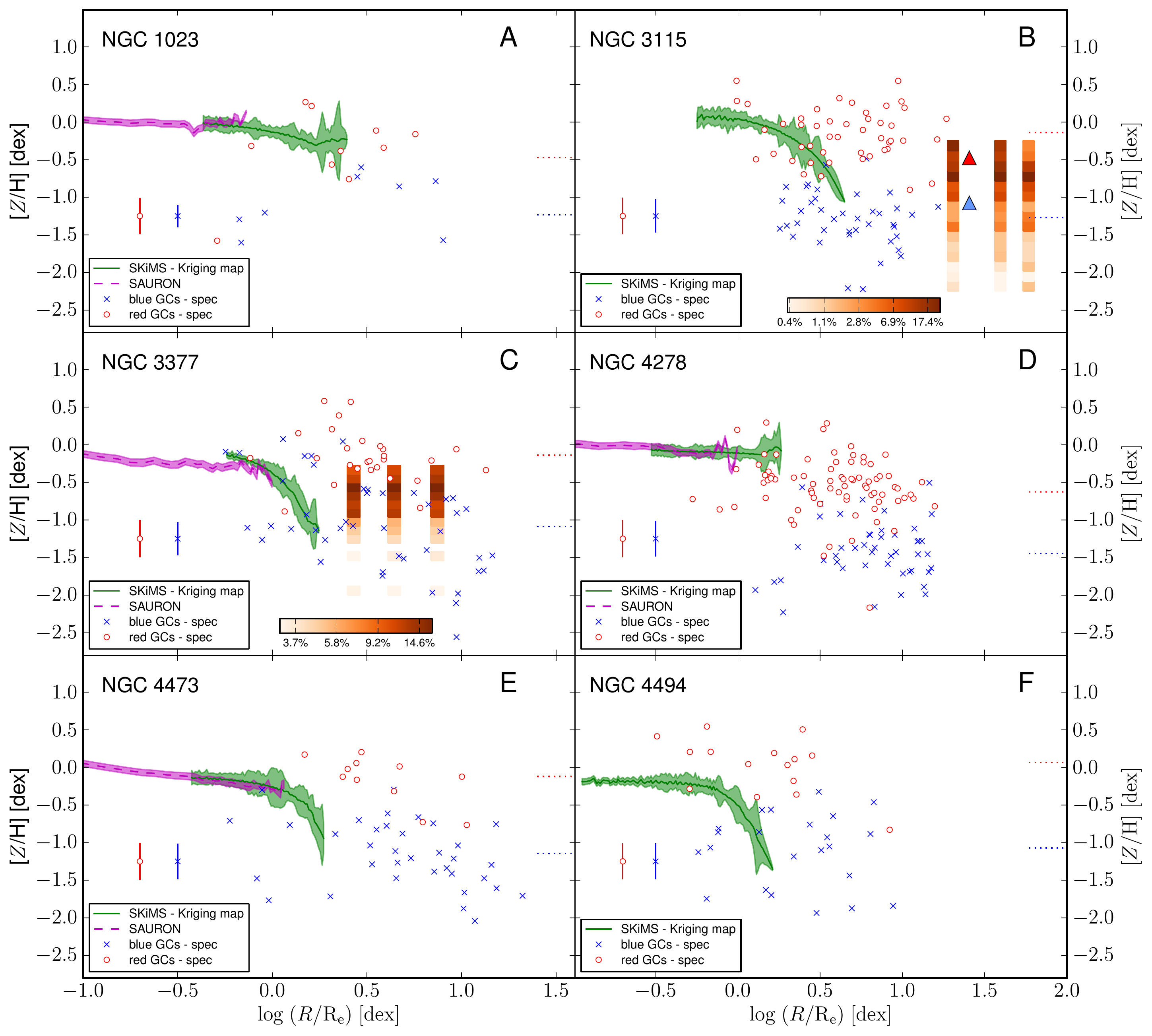}
		\end{center}
	    \caption[]{
	    		Metallicity profiles as in Figure \ref{fig:A1}, but for the 6 
	    		low-mass galaxies in the sample. 
	    		For NGC~3115 (panel B) and NGC~3377 (panel C) the density plots 
	    		show the photometric stellar metallicity distribution functions 
	    		(MDF) from \citet{Peacock15} and \citet{Harris07a}, respectively. 
	    		The radial position of the density plots corresponds to the 
	    		centre of the observed fields. 
	    		For better readability, in both cases the radial extension of 
	    		the density plots does not represent the actual size of the 
	    		observed field. 
	    		These density plots are colour-coded according to the colour 
	    		bars in the same panels.  
	    		For NGC~3115 (panel B) we also show the photometric metallicities 
	    		of the halo metal-rich and metal-poor stellar populations as 
	    		found by \citet{Elson97} as red and blue triangles, respectively. 
	    	    }
    \label{fig:A2}
  \end{figure*}
}

\newcommand{\placefigB}{
	\begin{figure*}
    	\begin{center}
			\includegraphics[width=2\columnwidth]{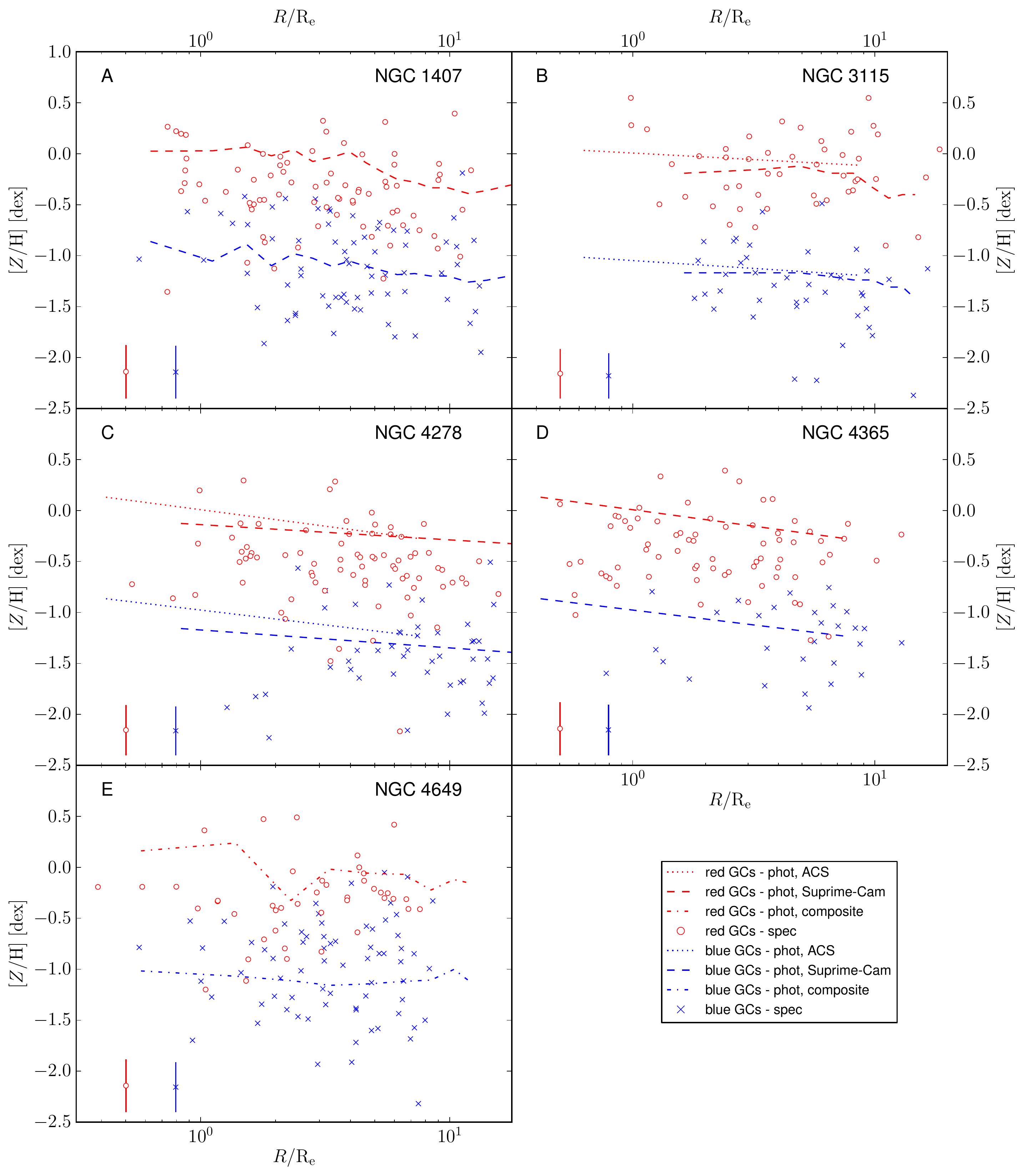}
		\end{center}
	    \caption[]{
	    		GC radial metallicity profiles. 
	    		Red circles and blue crosses show the spectroscopic 
	    		metallicities for the rGCs and bGCs, respectively. 
	    		The blue and red points with error bars show the average 
	    		uncertainties on the rGC and bGC spectroscopic metallicities, 
	    		respectively. 
	    		All the points are plotted at their elliptical galactocentric 
	    		radius scaled by $\rm{R_{e}}$. 
	    		In all the panels the red and blue lines show the literature 
	    		photometric metallicity profiles of rGCs and bGCs, 
	    		respectively. 
	    	    }
    \label{fig:B}
  \end{figure*}
}

\newcommand{\placefigC}{
	\begin{figure*}
    	\begin{center}
			\includegraphics[width=2\columnwidth]{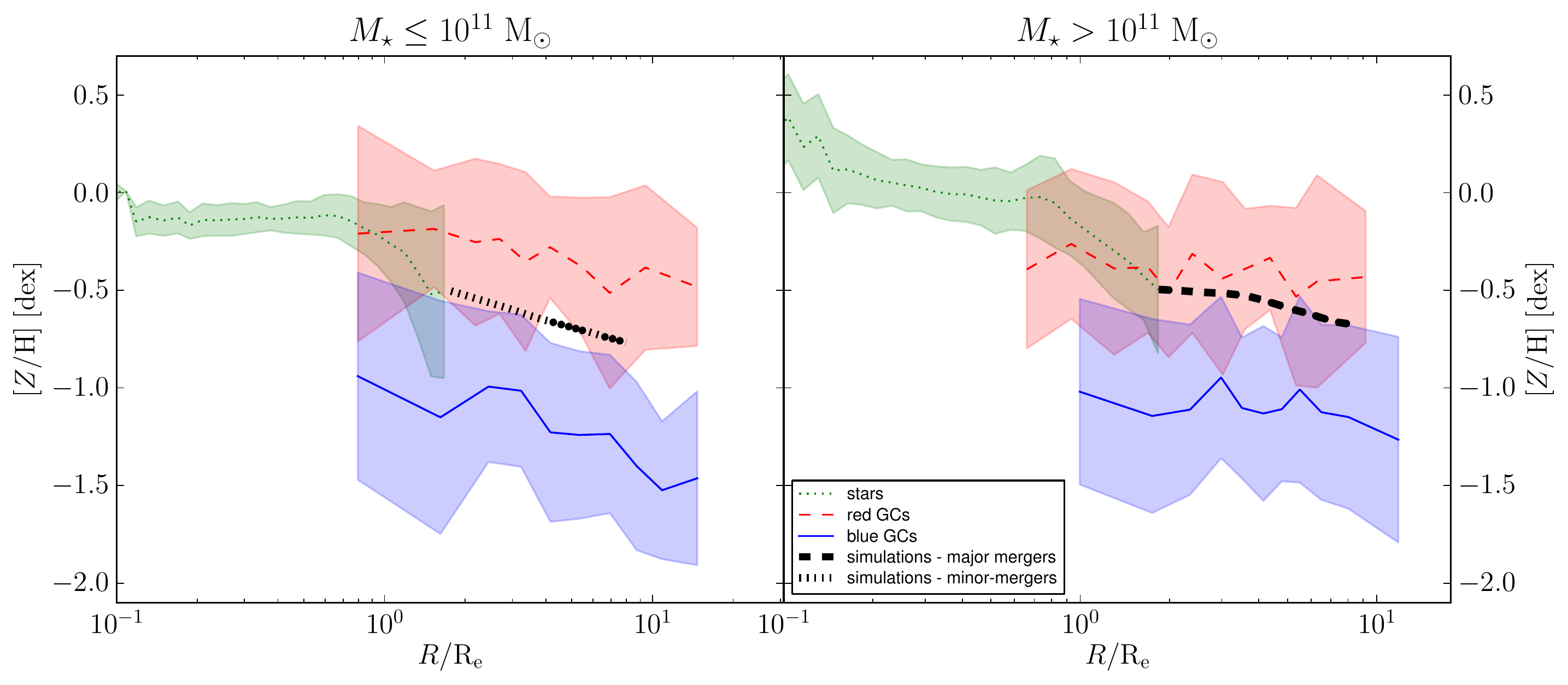}
		\end{center}
	    \caption[]{
	    		Combined stellar, rGC and bGC radial metallicity profiles for 
	    		the low-mass (\textit{left} panel) and high-mass (\textit{right} 
	    		panel) galaxies in our sample. 
	    		In both panels, the combined stellar, rGC and bGC metallicity 
	    		profiles are presented as a green dotted, a red dashed and a 
	    		blue solid line, respectively. 
	    		The green curves have been obtained combining SKiMS and 
	    		\sauron/\atlas metallicity profiles.
	    		The green, red and blue shaded region represent the $1\sigma$ 
	    		confidence limits in the stellar, rGC and bGC cumulated 
	    		metallicity profiles, respectively. 
	    		In the stellar case, the confidence intervals are dominated by 
	    		the scatter among the galaxies. 
	    		In the rGC and bGC cases, the confidence intervals are 
	    		dominated by the scatter among the GC metallicities within the 
	    		same galaxy. 
	    		The black dotted line in the \textit{left} panel and the 
	    		black dashed line in the \textit{right} panel show the 
	    		stellar metallicity profiles for the model M215 and M305 of 
	    		\citet{Hirschmann15}, respectively. 
	    		Such models represent two extreme formation 
	    		histories (i.e. M215 experienced only minor mergers while M305 
	    		has had a recent major merger). 
	    		These simulation stellar metallicity profiles have been scaled 
	    		in $[Z/H]$ to match with the photometrically observed 
	    		metallicity profiles of the high-mass sample in 
	    		\citet{LaBarbera12}. 
	    	    }
    \label{fig:C}
  \end{figure*}
}

\newcommand{\placefigE}{
	\begin{figure*}
    	\begin{center}
			\includegraphics[width=2\columnwidth]{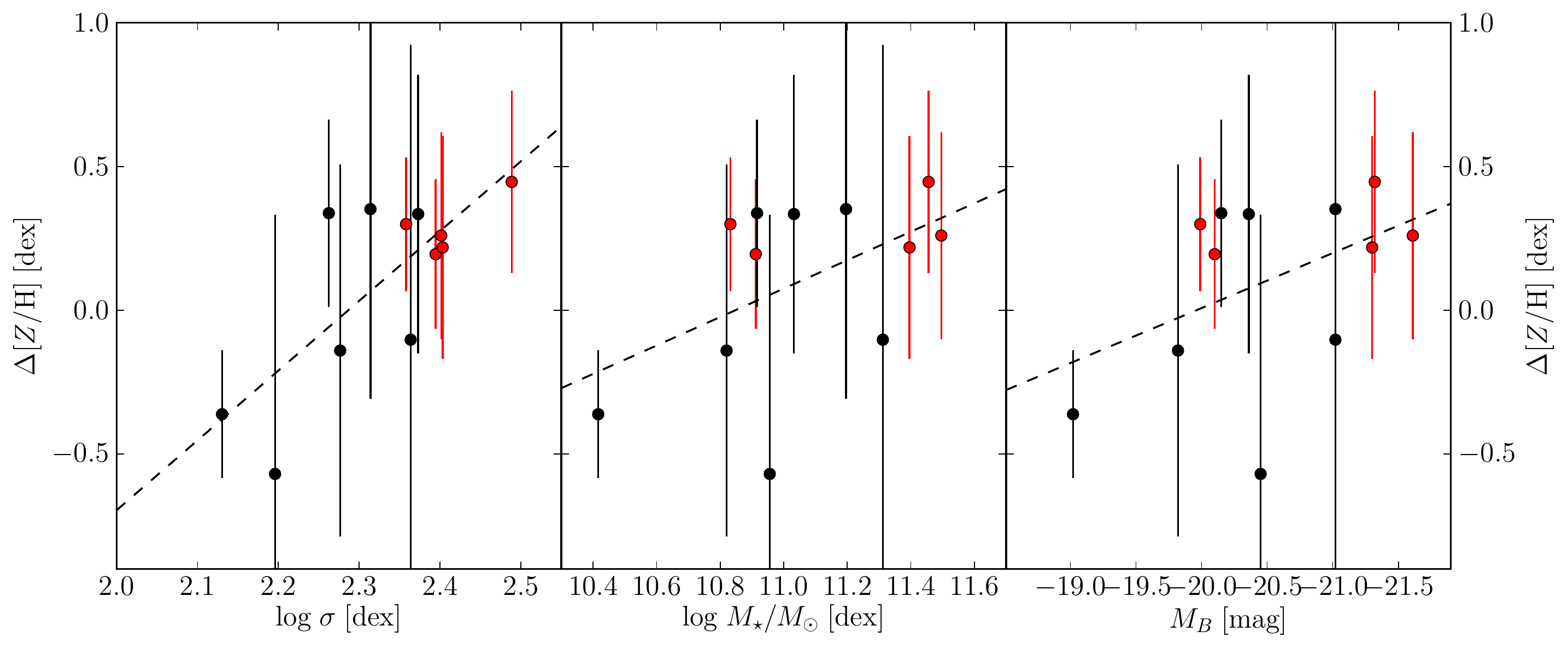}
		\end{center}
	    \caption[]{
	    		Difference between stellar and rGC metallicities (i.e. 
	    		$[Z/H]_{\rm{star}}-[Z/H]_{\rm{rGC}}$) versus various galaxy mass 
	    		proxies. 
	    		The stellar metallicities values are obtained at $1~\rm{R_{e}}$. 
	    		For 5 galaxies we have been able to extrapolate the rGC 
	    		metallicity to $1~\rm{R_{e}}$ from a linear fit of their radial 
	    		metallicity distribution. 
	    		These galaxies are presented as red points. 
	    		For the other galaxies, the rGC metallicities are obtained as 
	    		the average metallicity of the whole rGC system. 
	    		These galaxies are presented as black points. 
	    		In the \textit{left} panel the metallicity differences are 
	    		plotted against the stellar central velocity dispersion, in the 
	    		\textit{central} panel against the total stellar mass and 
	    		in the \textit{right} panel against the total $B$-band magnitude 
	    		of the host galaxy. 
	    		The dashed lines are the best-fitting linear functions in the 
	    		three cases. 
	    		A mild correlation between the stellar-rGC metallicity 
	    		difference and galaxy mass is seen. 
	    		}
    \label{fig:E}
  \end{figure*}
}

\newcommand{\placefigF}{
	\begin{figure*}
    	\begin{center}
			\includegraphics[width=2\columnwidth]{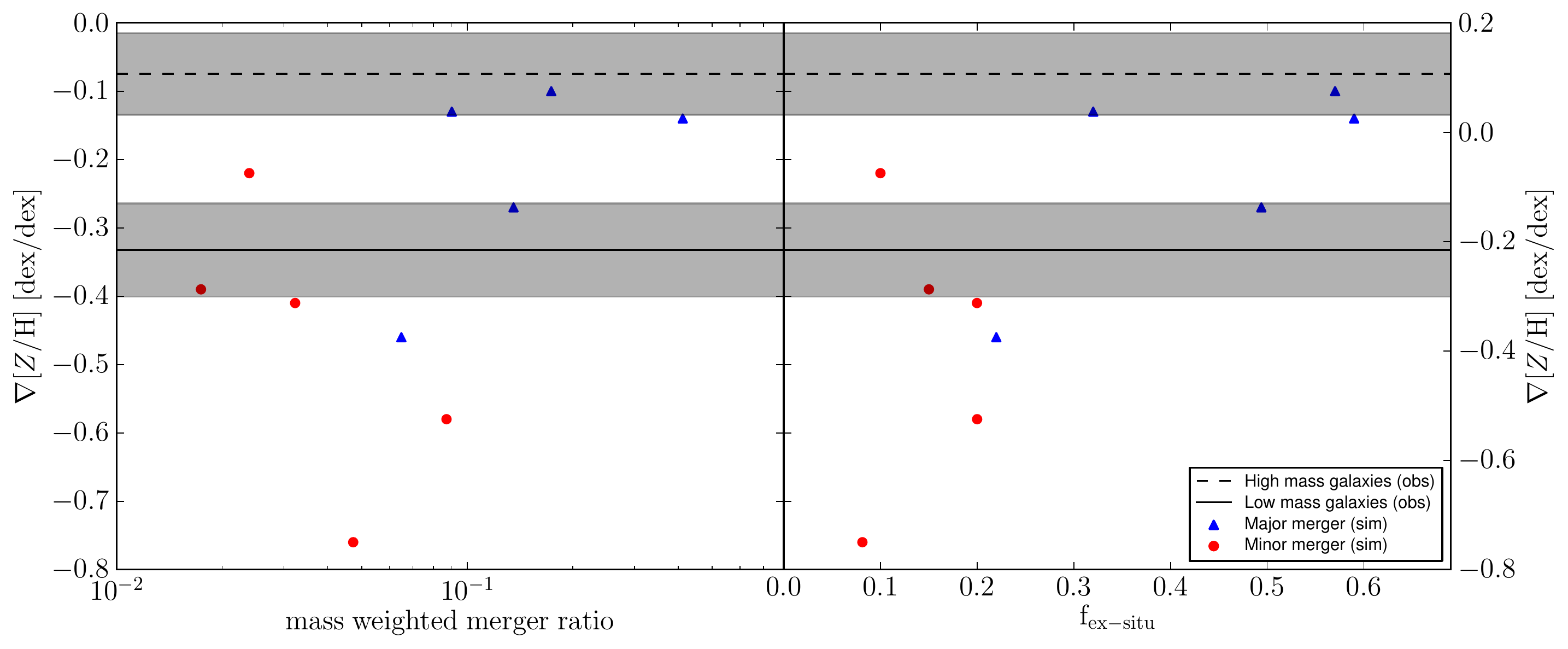}
		\end{center}
	    \caption[]{
	    		Outer metallicity gradients against mass weighted merger ratio 
	    		and ex-situ stellar fraction. 
	    		In the \textit{left} panel, the stellar metallicity gradients for  
	    		\citet{Hirschmann15} simulated galaxies in the wind feedback 
	    		models at $z=0$ are plotted against the mass weighted merger 
	    		ratio. 
	    		These gradients are measured in the radial range $2\leq R\leq
	    		6~\rm{R_{e}}$. 
	    		In the \textit{right} panel, the same stellar metallicity 
	    		gradients are plotted against the total fraction of ex-situ 
	    		stars in the simulations. 
	    		In both panels the red circles are for 
	    		minor merger dominated formation and blue triangles for 
	    		major merger dominated formation. 
	    		The solid and dashed black lines show the average metallicity 
	    		gradients for rGCs and bGCs obtained in this work for the low 
	    		and high galaxy mass bins, respectively. 
	    		These gradients are obtained in the whole available radial range 
	    		for our GCs. 
	    		Qualitatively, the GC metallicity gradient in our high mass 
	    		galaxies is similar to those in simulated galaxies with a high 
	    		fraction of ex-situ stars accreted via major mergers. 
	    		Our low mass galaxy GC metallicity gradient is similar to those 
	    		in simulated galaxies with a lower fraction of ex-situ stars and 
	    		more influence from minor mergers.  
	    		}
    \label{fig:F}
  \end{figure*}
}

\newcommand{\placefigAppendixGalaxy}{
	\begin{figure}		
    	\begin{center}
			\includegraphics[width=\columnwidth]{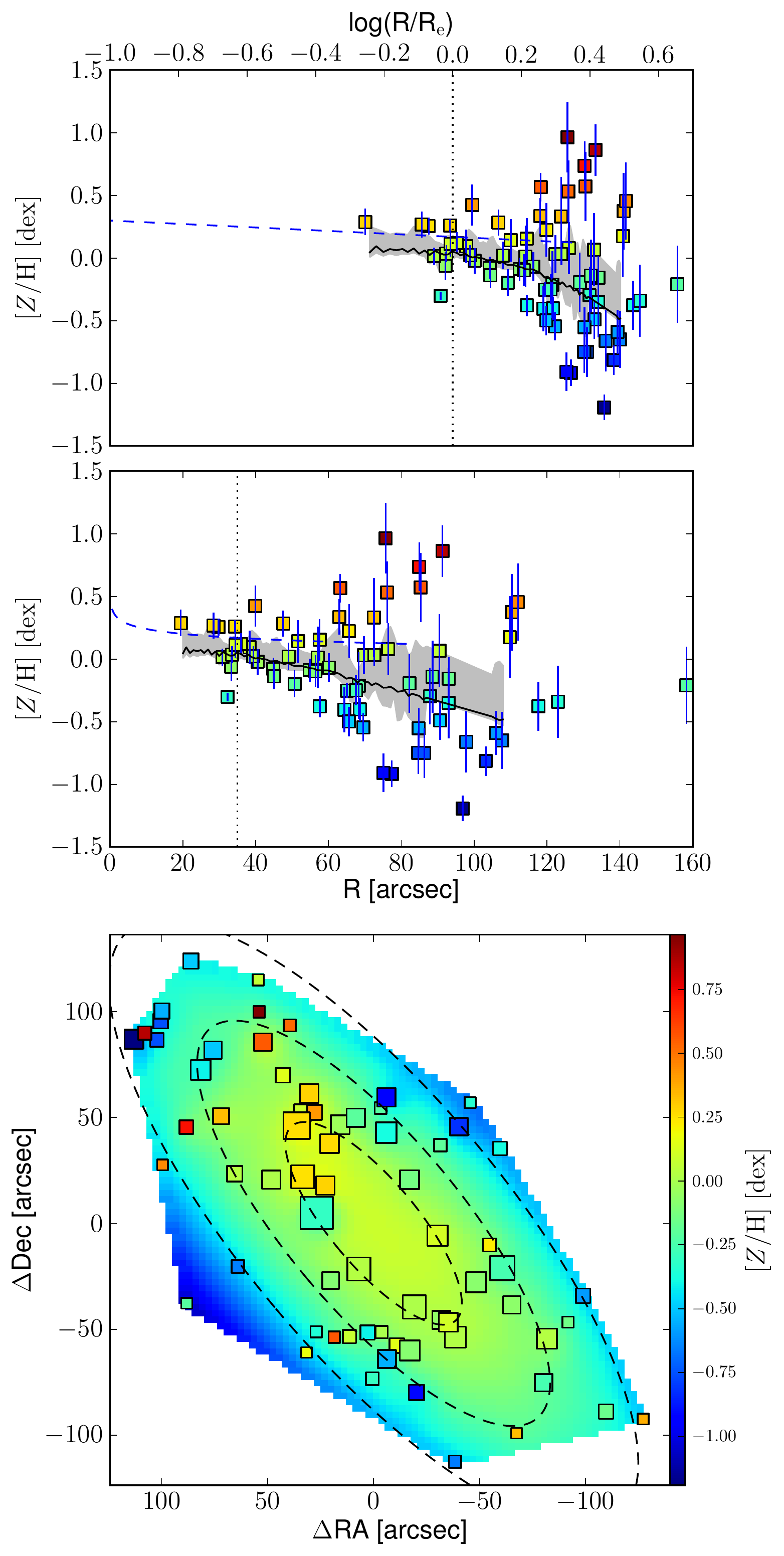}
		\end{center}
	    \caption[]{
	    1D metallicity profiles and 2D metallicity map of NGC~3115. 
	    The \textit{upper} panel shows the metallicity profile extracted from 
	    the kriging map (black solid line) versus the galactocentric radius in 
	    logarithmic space. 
	    The \textit{central} panel shows the same profile, but on a linear 
	    radial scale. 
	    In both these panels, the measured data points are shown as squares 
	    colour coded according to their metallicity. 
	    The black dotted vertical line shows the radius corresponding to 
	    $1~\rm{R_{e}}$. 
	    The blue dashed line shows the metallicity profile along the major 
	    axis as measured by \citet{Norris06}. 
	    The \textit{lower} panel shows the 2D metallicity map from kriging 
	    and the measured data points colour coded according to their metallicity 
	    values. 
	    The size of each point is inversely proportional to its uncertainty. 
	    The black dashed lines show the isophotes at $1$, $2$ and $3\rm{R_{e}}$, 
	    with ellipticity and PA from table 2 in \citet{Brodie14}.  
	    The data points, radial profiles and 2D metallicity map have been 
	    corrected with the empirical relation presented in equation 7 of 
	    \citet{Pastorello14}. 
	 	}
    \label{fig:AppA2}
  \end{figure}
}

\title[Stellar and globular cluster metallicities]{The SLUGGS survey: Combining 
		stellar and globular cluster metallicities in the outer regions of 
		early-type galaxies}

\author[Pastorello et al.]	
  {Nicola~Pastorello$^{1}$\thanks{email: npastorello@swin.edu.au},  
  {Duncan~A.~Forbes$^{1}$},
  {Christopher~Usher$^{1}$},
  {Jean~P.~Brodie$^{2}$},
\newauthor
  {Aaron~J.~Romanowsky$^{2,3}$},
  {Jay~Strader$^{4}$},
  {Lee~R.~Spitler$^{5, 6}$},
  {Adebusola~B.~Alabi$^{1}$},
\newauthor
  {Caroline~Foster$^{6}$}, 
  {Zachary~G.~Jennings$^{2}$},
  {Sreeja~S.~Kartha$^{1}$},
  {Vincenzo~Pota$^{2}$}
  \\
   $^1$Centre for Astrophysics \& Supercomputing, Swinburne University, Hawthorn 
   VIC 3122, Australia
  \\ $^2$University of California Observatories, 1156 High Street, Santa Cruz, 
  CA 95064, USA
  \\ $^3$Department of Physics and Astronomy, San Jos\'e State University, One 
  Washington Square, San Jos\'e, CA 95192, USA
  \\ $^4$Department of Physics and Astronomy, Michigan State University, East 
  Lansing, MI 48824, USA
  \\ $^5$Department of Physics and Astronomy, Faculty of Science and 
  Engineering, Macquarie University, Sydney, NSW 2109, Australia
  \\ $^6$Australian Astronomical Observatory, PO Box 915, North Ryde, NSW 1670, 
  Australia}

\begin{document}

\date{13$^{\rm{th}}$ January 2015}

\maketitle

\begin{abstract}
	The outer halo regions of early-type galaxies carry key information about 
	their past accretion history. 
	However, spectroscopically probing the stellar component at such 
	galactocentric radii is still challenging. 
	Using Keck/DEIMOS, we have been able to measure 
	the metallicities of the stellar and globular cluster components 
	in 12 early-type galaxies out to more than $10~\rm{R_{e}}$. 
	We find similar metallicity gradients for the metal-poor and metal-rich 
	globular cluster subpopulations, suggesting a common formation process for 
	the two subpopulations. 
	This is in conflict with most current theoretical predictions, where the 
	metal-poor globular clusters are thought to be purely accreted and 
	metal-rich globular clusters mostly formed in-situ. 
	Moreover, we find that the globular cluster metallicity gradients show a 
	trend with galaxy mass, being steeper in lower-mass galaxies than in 
	higher-mass galaxies. 
	This is similar to what we find for the outermost galaxy stars and 
	suggests a more active accretion history, with a larger role played 
	by major mergers, in the most massive galaxies. 
	This conclusion is qualitatively consistent with expectations from 
	two-phase galaxy assembly models. 
	% %
	Finally, we find that the small difference in metallicity between 
	galaxy stars and metal-rich globular clusters at $1~\rm{R_{e}}$ may 
	correlate with galaxy mass. 
	The origin of this difference is not currently clear. 

\end{abstract}


%%%%%%%%%%%%%%%%%%%%%%%
%												%
%  1. Introduction						%
%  \label{sec:introduction}		%
%												%
%%%%%%%%%%%%%%%%%%%%%%%
\section{Introduction}
\label{sec:introduction}

	The currently most popular view of massive early-type galaxy (ETG) formation 
	is the ``two-phase'' scenario. 
	The first phase consists of an early ($z>2$) dissipative collapse of gas 
	that forms the bulk of present-day galaxy stars. 
	The formation of these ``in-situ'' formed stars is driven by gas-rich 
	mergers and/or the gravitational in-fall of gas \citep{deLucia07, Dekel09b, 
	Zolotov09, Khochfar09b, Font11, Cooper15}. 
	The second phase, by contrast, involves the accretion of stars formed 
	``ex-situ'' (i.e. in external satellites), generally during gas-poor minor 
	merger events. 
	Simulations have found that such stars are mostly accreted in the outer 
	regions of the galaxy at $z < 2$ \citep{Naab09b, Oser10, Font11, Lackner12,
	Navarro-Gonzalez13}.
	In massive galaxies, an accretion of stars during major mergers events 
	is predicted to strongly reduce the stellar specific angular momentum 
	\citet{Khochfar11b}. 
	As a consequence, massive slow rotators are thought to have experienced 
	more major mergers than the fast rotators with similar mass. 

	We acknowledge that the separation of galaxy formation into two 
	phases might be somewhat artificial and too simplistic. 
	This is because mergers happen continuously, with the merger rate and gas 
	fraction decreasing steadily in time. 
	Therefore, the separation of galaxy formation in two phases at	$z\approx2$ 
	may be seen as the epoch that roughly denotes a shift from frequent gas-rich 
	accretion episodes to more discrete gas-poor merger events.  
	While the study of the bright central regions of ETGs has provided strong 
	constraints on their early formation phase, only the exploration of the 
	outer regions of galaxies will help to understand the second accretion 
	phase. 
	In addition, more than 90 per cent of the total mass and angular momentum, 
	and half of the stellar mass lie outside the half-light radius 
	($\rm{R_{e}}$). 
	Therefore, most of the information about a galaxy's formation and evolution 
	is encoded in its outer regions. 

	The two phase formation scenario makes prediction for the stellar 
	metallicity ($[Z/\rm{H}]$) gradients in ETGs. 
	The different phases of formation are expected to produce different 
	metallicity profiles for the ``in-situ'' and ``ex-situ'' components. 
	For instance, the first dissipative phase is supposed to create a metal-rich 
	``in-situ'' component in the inner regions, where the density is higher  
	and the stellar chemical evolution is faster \citep{Chiosi02, Zolotov10}. 
	This ``in-situ'' component will also reveal a steep radial metallicity  
	gradient, because of the rapidly decreasing local density with 
	galactocentric radius \citep{Kobayashi04, Pipino10}.

	On the other hand, the ``ex-situ'' stars form mostly in lower-mass, 
	relatively metal-poor, external systems. 
	In present-day ETGs, these stars contribute to lowering the metallicity in 
	the regions where they have been accreted, i.e., at large galactocentric 
	radii. 
	These outermost regions are expected to show generally shallow metallicity 
	gradients because of the dynamical mixing of pre-existing and accreted 
	stars \citep{Kobayashi04, diMatteo09a, Cooper10, Font11}. 

	In galaxies that experienced accretion, a transition 
	between the ``in-situ'' metal-rich and the ``ex-situ'' metal-poor dominated 
	regions is expected. 
	The radial position of such a transition, which marks where the accreted 
	component starts dominate, is still uncertain. 
	For example, the simulations of \citet{Hirschmann15} show that this 
	transition radius could lie from 4 to more than 8 $\rm{R_{e}}$.
	Because of their different accretion histories, the metallicity profiles of 
	low mass galaxies are expected to be systematically different from those of 
	their high mass counterparts. 

	Finally, major merger events in massive galaxies will weaken any 
	pre-existing stellar metallicity gradient \citep{White78, White80, 
	Kobayashi04, diMatteo09a, Rupke10, Navarro-Gonzalez13, Hirschmann15}. 
	This is because as the mass of the two galaxies involved in the merger are 
	similar, their stars will also show similar metallicities but will be mixed 
	at all galactocentric radii in the final galaxy. 

	To date, just a handful of studies have spectroscopically 
	probed the stellar metallicity at large radii, and most of them just in the 
	high-mass galaxy range (i.e. $M_{\star}>10^{11}~\rm{M_{\odot}}$, 
	\citealt{Coccato10, Greene12, Mihos13}).  
	An exception to this is \citet{Pastorello14}, where the stellar metallicity 
	of the outer region ($1<R\leq2.5\rm{R_e}$) has been spectroscopically 
	measured also in lower-mass galaxies (i.e. with stellar mass down to 
	almost $10^{10}~\rm{M_{\odot}}$). 
	In this work, the outer metallicity gradient was found to correlate with 
	galaxy mass, becoming steeper in low-mass galaxies. 
	This may confirm that high-mass galaxies accrete more, and higher-mass, 
	satellites than their low-mass counterparts. 
	However, the study of \citet{Pastorello14} did not reach the galactocentric 
	radii where the accreted component is expected to dominate. 

	For a small number of very nearby galaxies, stellar metallicity studies at 
	very large radii have been based on the photometric analysis of resolved 
	red-giant branch (RGB) stars \citep{Elson97, Harris07a,Harris07b, Bird15, 
	Rejkuba14, Peacock15}. 
	A transition between possible ``in-situ'' and ``ex-situ'' stars 
	has been noticed around $12~\rm{R_{e}}$ in NGC~3379 \citep{Harris07b}. 
	In a spectroscopic study of M31 RGB stars, \citet{Kalirai06} found a 
	transition in the stellar metallicity at $\approx30~\rm{kpc}$ from the 
	galaxy centre. 
	As this method requires resolved stars, it is limited to the local Universe, 
	and just a handful of galaxies have been studied in this way. 
	
	A way to overcome such issues and explore a higher number of galaxies out to 
	large radii is to use a different tracer of the galaxy stellar populations. 
	In particular, globular clusters (GCs) are a viable way to explore the 
	metallicity at large galactocentric radii, since they are bright and dense 
	stellar systems. 
	For this reason, their study can extend beyond the detectable galaxy 
	stellar light, and their metallicities can be measured out to more than 
	$10~\rm{R_{e}}$ \citep{Usher12}. 
	Since GCs are generally very old (i.e. age $>10~\rm{Gyr}$), they have 
	survived the evolutionary processes that formed their present-day host 
	galaxies. 

	Almost all the GC systems of galaxies with $M_{\star} \gtrsim10^{10}~
	\rm{M_{\odot}}$ present a bimodality in colour and metallicity, 
	with blue GCs (bGCs) showing lower metallicities than red GCs (rGCs) 
	(\citealt{Strader06, Brodie06, Peng06}, and references therein). 
	The origin of this bimodality is still unclear, but it is commonly 
	believed that it is intrinsically linked with the host galaxy's formation 
	history (e.g. \citealt{Strader05, Brodie06}). 
	In general, rGCs seem to be linked with the host galaxy bulge's stellar 
	population, having similar spatial distribution, kinematics and metallicity, 
	while bGCs seem to share several properties with the galaxy halo component 
	(e.g., \citealt{Faifer11, Strader11, Arnold11, Forbes11, Forbes12, Pota13,
	Peacock15, Kartha14, Pota15}). 
		
	In this paper we explore galaxy stellar populations at large radii 
	using both stellar and GC metallicities. 
	In particular, we find that the steepness of the GC metallicity radial 
	profile is higher in low mass galaxies, qualitatively similar to 
	that between stars and galaxy mass found by \citet{Pastorello14}. 
	Furthermore, we measure the metallicity offset between the galaxy stars 
	in the innermost regions and the rGCs. 
	We find that this offset correlates with the galaxy mass, in the sense that 
	higher-mass galaxies present larger offsets. 

	The paper is structured as follows. 
	In Section \ref{sec:data}, we present the data and the methods we use to 
	obtain the metallicity of both stars and GCs. 
	In particular, in Section \ref{sec:data_comparison} we compare our 
	spectroscopic metallicities with photometric metallicity profiles from the 
	literature in order to assess the possible selection bias for our sample of 
	spectroscopic targets. 
	Section \ref{sec:results} presents the results we obtain from the radial 
	metallicity profiles for both the stellar and the GC components. 
	In Section \ref{sec:discussion} we present our findings and discuss how they 
	compare with the current theoretical predictions and literature results. 
	Finally, in Section \ref{sec:conclusions} we provide a summary of the 
	results. 
	In addition, in Appendix A we present an updated metallicity map for the 
	galaxy NGC~3115, reflecting new extra data points obtained since the 
	publication of the original in \citet{Pastorello14}.

%%%%%%%%%%%%%%%%
%								%
%  2. Data acquisition and analysis				%
%  \label{sec:data}	%
%								%
%%%%%%%%%%%%%%%%

\section{Data acquisition and analysis}
\label{sec:data}

	\placetabA

	\subsection{The SLUGGS sample}
		In this work we present the metallicity radial profiles of both the 
		stellar and the GC components for 12 galaxies from the 
		ongoing SAGES Legacy Unifying Globulars and GalaxieS 
		(SLUGGS\footnote{http://sluggs.swin.edu.au}) survey \citep{Brodie14}. 
		The survey sample includes 25 early-type galaxies (i.e. ellipticals and 
		lenticulars) within a distance $D<30~\rm{Mpc}$. 
		One of the aims of this survey is the study of the stellar populations 
		in both GCs and field stars. 
		We select the sub-set of 12 SLUGGS galaxies for which the metallicities 
		of at least 15 spectroscopically confirmed GCs have been measured. 

		The SLUGGS dataset presented in this paper has been obtained over 
		9 years and 25 observing runs. 
		The GC photometry is from \citet{Usher12} for all the galaxies, except 
		NGC~4473 \citep{Alabi15}, NGC~4649 \citep{Pota15} and NGC~1023 (this 
		work). 
		Table \ref{tab:A} presents the adopted parameters for the galaxies in 
		this paper. 

		\subsection{DEIMOS spectra}
		In order to reduce the DEIMOS spectra, we adopt a modified version of 
		the IDL \texttt{spec2D} pipeline \citep{Cooper12, Newman13}, as 
		presented in \citet{Arnold14}. 
		In multi-slit mode, from a single DEIMOS slit, it is possible to obtain 
		the spectra of both the target object and the background light. 
		While the former is used to study the GC stellar population properties, 
		the latter includes light from both the galaxy's stars and the night 
		sky. 
		Adopting the ``Stellar Kinematics with Multiple Slits'' (SKiMS) 
		technique \citep{Norris08, Proctor09, Foster09, Pastorello14, Arnold14, 
		Foster15}, it is possible to separate the two components in the 
		background spectra and obtain an individual spectrum of the GC and the 
		galaxy starlight at each slit location. 

	\subsection{Stellar metallicities}\label{sec:stellarMet}
		The SKiMS metallicity data obtained from the CaT index have been 
		corrected following the prescriptions in section 3.4 of 
		\citet{Pastorello14}. 
		In particular, in the galaxies for which \sauron\ or \atlas\ data are 
		available, we match the radial profiles within 1~$\rm{R_{e}}$ in order 
		to minimize the metallicity offset. 
		The \sauron\ metallicity data are the same used in \citet{Pastorello14}, 
		to which we add the \atlas\ metallicity data for NGC~4365 kindly 
		provided by Harald Kuntschner (Kuntschner et al., in prep.). 
		When the galaxy metallicity is not available from the \sauron\ and 
		\atlas\ surveys, we adopted the \citet{Pastorello14} empirical 
		correction relation:
		\begin{eqnarray}
			\Delta [Z/\rm{H}] = 3.78 \cdot \log \sigma_{\rm{0}} - 8.08
		\end{eqnarray}
		where the values for the central velocity dispersion $\sigma_{\rm{0}}$ 
		are those in Table \ref{tab:A}. 

		From these metallicity values, we obtained 2D maps of the galaxy 
		metallicity using the kriging technique (see \citealt{Pastorello14} for 
		details). 
		Kriging maps take into consideration the non-optimal spatial sampling of 
		the available datapoints. 
		In this way we are able to obtain pseudo-IFU maps out to several 
		effective radii and derive radial metallicity profiles. 
		For this purpose, we are adopting the photometric parameters in Table 
		\ref{tab:A}. 

		Throughout the paper, the galactocentric radius is expressed in terms of 
		circularized radius, defined as: 
		\begin{eqnarray}
			R = \sqrt{qx^{2}+\frac{y^2}{q}}
		\end{eqnarray}
		where $q$ is the axial ratio and $x,y$ are the coordinates along the 
		major and the minor axes of the galaxy, respectively.
		For consistency, the same definition of radius is also used to convert 
		the galactocentric radii of literature metallicity values, when 
		these are given on different scales. 

	\subsection{GC metallicities}
		The metallicities of the GCs presented in this work are mostly 
		those presented in \citet{Usher12}, \citet{Usher13} and \citet{Usher15}. 
		These are supplemented with new data obtained from recent Keck 
		observations
		reduced and analysed using the methods described in \citet{Usher12}. 
		In particular, we add new data for NGC~1023, NGC~1407 and NGC~3115. 
		From the total sample, we select the galaxies with more than 15 
		spectroscopically confirmed GCs with $S/N > 10~\rm{\AA}$ in the 
		wavelength range from $8400$ to $8500~\rm{\AA}$. 

		In order to separate the blue and the red GC subpopulations we 
		adopt a split in $(g-i)$ colours as defined in Table \ref{tab:A}. 
		Such colour values are obtained from the best fitting bimodal 
		colour distributions calculated with the Gaussian Mixture Modelling (GMM)
		code of \citet{Muratov10}, as in \citet{Usher12}. 
		The whole sample of GC CaT and metallicities used in this paper are 
		presented in Table \ref{tab:AppData}. 

	\subsection{Stellar and GC metallicity scales}

		The stellar and the GC metallicities used in this work are defined on 
		slightly different metallicity scales. 	
		While \sauron/\atlas\ metallicities are obtained from Lick indices and 
		stellar population models from \citet{Schiavon07}, the 
		\citet{Pastorello14} stellar metallicities are from the CaT index and 
		the SSP models of \citet{Vazdekis03}. 
		In addition, an empirical correction is applied to the 
		\citet{Pastorello14} metallicities to correct for an offset with 
		\sauron/\atlas\ metallicities in the overlapping radial regions, which is 
		likely driven by the IMF. 
		GC metallicities are also obtained from the CaT index and the SSP models 
		of \citet{Vazdekis03} (for details, see \citealt{Usher12}). 
		Any differences in the metallicity scale between stellar and GC 
		metallicities may, in principle, lead to systematic offsets between the 
		two components. 

	\subsection{Comparison between spectroscopic and photometric GC metallicity 
				radial profiles}\label{sec:data_comparison}

		Our sample of GC spectroscopic metallicity may suffer from a selection 
		bias at low metallicities, given that this selection is based on 
		the detection of metallicity-dependent absorption lines. 
		However, the GC metallicity of several galaxies in our sample has 
		already been studied photometrically. 
	  	For these galaxies we compare our spectroscopic GC metallicities with 
	  	the photometric GC metallicities from the literature, in order to check 
	  	whether or not our sample is representative of the whole GC population.  
  		Most GCs are predominately old, so photometric colours are a good proxy 
  		for metallicity \citep{Puzia05, Strader05, Spitler08},
  		although the relationship is non-linear and seems to vary 
  		galaxy-to-galaxy \citep{Usher15}.

		In the case of NGC~1407, \citet{Forbes11} presented ($g-i$) radial 
		trends of the red and blue GCs from Subaru/Suprime-Cam. 
		We convert these colours into metallicities following equation 10 in 
		\citet{Usher12}. 
		In their work, \citet{Forbes11} found that the colour gradients of both 
		rGCs and bGCs are steeper for $R<10~\rm{arcmin}$ and flatten in the 
		outer regions. 
		They were concerned that the flattening of the rGC colour gradient in 
		the outer regions may be a consequence of strong contamination (more 
		than 50 per cent) outside of $10~\rm{arcmin}$ from the galaxy centre. 
		We reduce the probability of this case, since we spectroscopically 
		confirm the presence of more than few rGCs out to almost $15~	
		\rm{arcmin}$ in this galaxy. 

		The two subpopulations of GCs in NGC~3115 have been studied by 
		\citet{Arnold11} and \citet{Jennings14}. 
		\citet{Arnold11} presented the ($g-i$) colour profiles obtained 
		from Suprime-Cam/Subaru photometry. 
		We extract these profiles from their figure 3 and convert them into 
		metallicity profiles using equation 10 in \citet{Usher12}. 
		\citet{Jennings14} presented the ($g-z$) colour profiles 
		obtained from \textit{HST}/ACS photometry. 
		To convert these colours into total metallicities, $[Z/\rm{H}]$, we adopt 
		equation B1 in \citet{Usher12}. 
		% %

		\citet{Usher13} presented the colour gradients of NGC~4278 rGCs and bGCs 
		from both \textit{HST}/ACS and Subaru/Suprime-Cam photometry. 
		From these, we obtain the ($g-z$) and ($V-I$) colour 
		gradients for both the GC subpopulations, which we convert into total 
		metallicities $[Z/\rm{H}]$ with equation B1 and B3 in \citet{Usher12}, 
		respectively. 

		The case of NGC~4365 is peculiar, since this galaxy hosts three, rather 
		than two, GC subpopulations \citep{Puzia02}. 
		\citet{Blom12a} presented the ($g-i$) colour profiles obtained 
		from Subaru/Suprime-Cam photometry for the three subpopulations. 
		For the purposes of a spectroscopic/photometric metallicity comparison, 
		we consider only the rGC and the bGC colour profiles in their work and 
		exclude the central ``green'' GCs. 
		We extract these colour profiles from their figure 16 and convert them 
		into metallicity profiles with equation 10 in \citet{Usher12}.

		Similarly, \citet{Pota15} present the ($g-z$) colour profiles 
		for both the rGC and bGC subpopulations in NGC~4649.  
		These colours are obtained from \textit{HST}/ACS, Subaru/Suprime-Cam and 
		CFHT/MegaCam photometry. 
		As per \citet{Jennings14}, we convert \citet{Pota15} colours into total 
		metallicities $[Z/\rm{H}]$ using equation B1 in \citet{Usher12}. 

		\placefigB
	
		In Figure \ref{fig:B} the comparison between photometric and 
		spectroscopic metallicities is presented for NGC~1407, NGC~3115, 
		NGC~4278, NGC~4365 and NGC~4649. 
		In all the galaxies, the bGC average spectroscopic metallicities are 
		generally consistent with the photometric metallicities. 
		We therefore exclude the possibility of a significant bias at 
		low-metallicity for our spectroscopic bGC sample. 
		However, the spectroscopic rGC metallicities are generally lower than 
		the photometric ones at the level of $\approx0.3~\rm{dex}$. 
		This may be a consequence of (1) non universal colour-metallicity 
		relations for GCs, as discussed in \citet{Usher15}, (2) the 
		spectroscopic sample not being representative of the GC subpopulation, 
		and/or (3) an imperfect classification of GCs into the two 
		subpopulations (which is based on the GC colours). 
		In this latter case, if a non negligible number of rGCs are, instead, 
		misclassified bGCs, then the overall rGC average spectroscopic 
		metallicity would be higher. 
		This would happen if the rGC and bGC colour distributions are not 
		symmetrical, with the distribution of bGCs showing a significant 
		extension to redder colours.
		However, we note that this explanation is purely speculative, since 
		the only models showing a possible asymmetry between the colour 
		distributions of the two subpopulations we are aware of show a possible 
		contamination of bGCs by rGCs and not the opposite \citep{Forte07}.
		In any case, the offset between photometric and spectroscopic rGC 
		metallicities is comparable to the uncertainties on the actual 
		spectroscopic measurements an the systematic uncertainties in the 
		colour-metallicity conversions. 
		We conclude that, at least in the galaxies for which we have a 
		high number of spectroscopically measured GC metallicities, the spectral 
		subsample is a reasonable representation of the true metallicity 
		distribution. 
		%

%%%%%%%%%%%%%%%%%%%%%
%											%
%  4. Results and Discussion						%
%  \label{sec:results}		%
%											%
%%%%%%%%%%%%%%%%%%%%%

\section{Results}
\label{sec:results}
	In order to disentangle the different behaviours of the stellar and GC 
	metallicities with galaxy mass, we split our sample into two different 
	stellar mass bins. 
	The radial metallicity profiles for the stellar and GC components of the 
	12 galaxies in this work are presented in Figure \ref{fig:A1} (for 
	the galaxies with $M_{\star} > 10^{11}~M_{\odot}$) and 
	Figure \ref{fig:A2} (for the galaxies with $M_{\star} \leq 
	10^{11}~M_{\odot}$). 

	Our $[Z/\rm{H}]$ stellar radial profiles are 
	extracted from the kriging maps presented in \citet{Pastorello14}. 
	In addition, for 7 galaxies we have obtained their inner radial 
	metallicity profiles from \sauron/\atlas\ maps. 
	While the former do not extend to the galaxy centre, the latter do 
	not extend to $1~R_{e}$ and are, thus, complementary to our data. 
	The stellar metallicity profiles, their correction to match with the 
	inner \sauron/\atlas\ profiles and the stellar gradient trends with galaxy 
	mass (i.e. steeper in lower mass galaxies) are discussed in 
	\citet{Pastorello14}. 

	\placefigAfirst
	\placefigAsecond

	\subsection{Halo stellar metallicity from resolved RGB stars}
		Here we compare our integrated stellar metallicities and GC 
		metallicities with those from resolved RGB stars from \textit{HST} 
		observations. 
		A caveat for these studies is that the RGB measurements are usually 
		incomplete at high metallicities, and therefore their values may 
		underestimate the mean metallicities of the real stellar population. 
		% %
		For two of our galaxies, NGC~3115 and NGC~3377, RGB metallicities are 
		available and we include these values in Figure \ref{fig:A2}. 

		\citet{Elson97} presented the RGB stellar metallicity for NGC~3115 
		measured from an \textit{HST}/WFPC2 field at $R\approx25~\rm{R_{e}}$. 
		She found two distinct stellar populations, with iron abundances of 
		$[\rm{Fe} / \rm{H}] \approx -0.7$ and $-1.3~\rm{dex}$, respectively. 
		In order to plot her results in our Figure \ref{fig:A2}, we  
		convert them with the \citet{Thomas03a} relation: 
		\begin{eqnarray}\label{ThomasRel}
			[Z/\rm{H}] = [\rm{Fe}/\rm{H}] + 0.93~[\alpha/\rm{Fe}]
		\end{eqnarray}
		where $[\alpha/\rm{Fe}]$ is the $\alpha$-element abundance. 
		\citet{Norris06} measured a radially constant 
		$[\alpha/\rm{Fe}]\approx0.17~\rm{dex}$ along the NGC~3115 major axis, 
		while they found a variable $\alpha$-element abundance along the minor 
		axis in the range $0.17\leq[\alpha/\rm{Fe}]\leq0.30~\rm{dex}$. 
		To deal with this variability, we assume $[\alpha/\rm{Fe}]=0.24~\rm{dex}$. 
		We note that the low-metallicity subpopulation presented by 
		\citet{Elson97} has a comparable number of stars to the 
		high-metallicity subpopulation. 
		This may be due to instrumental calibration issues, as later 
		pointed out by \citet{Kundu98}.

		More recently, for the same galaxy, \citet{Peacock15} obtained RGB 
		metallicities for three different \textit{HST}/ACS and 
		\textit{HST}/WFPC3 fields. 
		These are also plotted in Figure \ref{fig:A2}. 
		We confirm that the bulk of their measurements are consistent with 
		a metallicity in between the red and the blue GC metallicities at the 
		same radii. 
		In addition, the low metallicity tails in the three fields extend into 
		the bGC metallicity domain. 
 		\citet{Peacock15} also found evidence for a low-metallicity 
		stellar population in their outermost peak (at $R=53.8~\rm{R_{e}}$), with 
		an average metallicity value (i.e. $[Z/\rm{H}]\approx-1.2~\rm{dex}$) similar 
		to that of \citet{Elson97}, albeit with the number of RGB stars 
		significantly lower than that presented in \citet{Elson97}. 
		Interestingly, this low-metallicity stellar population has an 
		average metallicity very similar to that of the bGCs (i.e. 
		$[Z/\rm{H}]=-1.27~\rm{dex}$). 

		Similarly, we add \citet{Harris07a} RGB stellar photometric 
		metallicities from \textit{HST}/ACS imaging in the outer regions of 
		NGC~3377. 
		The targeted three contiguous fields, all yielding to a consistent 
		metallicity of $[Z/\rm{H}]\approx-0.6~\rm{dex}$. 
		This is in conflict with the general trend of our SKiMS metallicities up 
		to $2~\rm{R_{e}}$, which instead show a steep decline down to 
		$[Z/\rm{H}]\approx-1.0~\rm{dex}$. 
		Since the \citet{Harris07a} fields probed only a small fraction of the 
		halo, a possible azimuthal variation or substructure of the RGB stellar 
		metallicity may explain the observed difference. 

		\placefigC

 	\subsection{GC spectroscopic metallicity gradients}\label{sec:results_ourresults}
		In Figure \ref{fig:C} we show the combined metallicity profiles for 
		both the stellar and the GC components for our two galaxy stellar 
		mass bins. 
		To obtain the combined stellar metallicity profiles in the two 
		mass bins, we stack the stellar metallicity profiles of the single 
		galaxies and derive the average profile (and the standard deviation) 
		within logarithmic radial bins of $0.05~\rm{dex}$. 
		The combined rGC (bGC) metallicity profiles are obtained stacking the 
		metallicities of all the rGCs (bGCs) in each galaxy mass bin. 
		We derive the average profile (and the standard deviation) within radial 
		bins with a constant number of elements (i.e. $N_{\rm{bin}}=20$).

		In Table \ref{tab:D} we present the metallicity gradients measured for 
		the rGC and bGC subpopulations in the two galaxy mass bins with 
		separation at $M_{\star} = 10^{11}~M_{\odot}$. 
		In the low mass galaxy bin, the rGC and bGC gradients are consistent 
		within the associated uncertainties. 
		Likewise, the rGC and bGC metallicity gradients in the high mass bin are 
		consistent with each other within the associated uncertainties, and are 
		shallower than the rGC and bGC metallicity gradients in the low mass 
		bin. 
		In particular, both rGC and bGC gradients show a strong difference 
		between the two mass bins (i.e. $1.7\sigma$ and $2.3\sigma$, 
		respectively). 
		Such gradients are measured in the whole available radial 
		range for the different GC subpopulations (we find that using a 
		restricted, common radial range does not change the conclusions).

		It is useful to remember that, in general, 3D metallicity profiles are 
		$\approx10\%$ steeper than those measured for projected galactocentric 
		radii \citep{Liu11}. 

		\placetabD

    \subsection{Comparison with metallicity gradients from simulations}\label{comparisonHirschmann}
    	From hydrodynamical high-resolution, cosmological zoom simulations, 
    	\citet{Hirschmann15} obtained stellar metallicity profiles out to 
    	large radii for 10 galaxies with stellar mass in the range $1.7\times
    	10^{11}<M_{\star}<3.7\times10^{11}~\rm{M_{\odot}}$. 
    	They found that when including stellar wind feedback the average stellar 
    	metallicity profile slope is consistent with the photometric 
    	observations by \citet{LaBarbera12}. 
    	Hence, we compare our metallicity gradient results with the predictions 
    	from only their	simulations including stellar wind feedback. 
    	In what follows we assume that the \citet{Hirschmann15} gradient 
    	predictions are predominately driven by the different accretion 
    	histories of the simulated galaxies, and not by the mass (which has a 
    	very narrow range).
    	We do not compare the absolute metallicity values of 
    	\citet{Hirschmann15}, although we do apply an offset in order to have 
    	the profiles match photometric observations from the literature. 
    	This is because the absolute value of stellar metallicity in 
    	simulations is sensitive to the subgrid baryonic physics.  
    	Therefore, it is not safe to directly compare our absolute values with 
    	those from simulations, while instead we can qualitatively compare the 
    	metallicity gradients (which depend mostly on the galaxy formation 
    	history).

    	In figure 4 of \citet{Hirschmann15}, the average metallicity profile is 
    	arbitrarily scaled in metallicity for visualization purposes (M. 
    	Hirschmann, private communication). 
    	We extract from this figure the average metallicity profile and offset 
    	it to match the observed profile for the high-mass galaxy sample in 
    	figure 8 of \citet{LaBarbera12}. 
    	In particular, we adopt the \citet{LaBarbera12} metallicity profile 
    	obtained with the \citet{Bruzual03} single stellar population models. 
    	In this way we obtain the vertical offset to apply to the 
    	\citet{Hirschmann15} profiles. 

    	In Figure \ref{fig:C} we plot two extreme examples of 
    	assembly history from the \citet{Hirschmann15} simulations with 
    	stellar wind feedback at $z=0$, representative of minor and major merger 
    	dominated evolutionary history, respectively. 
    	In particular, in the \textit{left} panel (i.e. our low-mass galaxy bin) 
    	we plot the stellar metallicity profile from the minor merger model 
    	M215, which is the model with the highest fraction of minor 
    	mergers and no past major merger event. 
    	In the \textit{right} panel (i.e. our high-mass galaxy bin) we plot the 
    	stellar metallicity profile from the major merger model M305. 
    	This is the model with the highest accreted mass fraction from 
    	major mergers. 
    	In particular, it includes a major merger event at $z\approx0$. 
    	In both cases, the vertical offsets applied to the simulated stellar 
    	metallicity profiles provide a good match with our spectroscopically 
    	measured stellar metallicity profiles at the common radius. 

    	\citet{Hirschmann15} provided the stellar metallicity gradients in the 
    	range $2<R<6~\rm{R_{e}}$. 
    	The stellar metallicity profile of M215 has a gradient of $-0.41~
    	\rm{dex/dex}$, consistent with that of bGCs in our low-mass bin 
    	(Table \ref{tab:D}). 
    	The stellar metallicity gradient of M305 has a gradient of $-0.13~
    	\rm{dex/dex}$, consistent with both the gradients of rGCs and bGCs in 
    	our high-mass bin (Table \ref{tab:D}). 

		The range of metallicity gradients in the minor merger 
		simulations by \citet{Hirschmann15} is $-0.22\leq \nabla [Z\rm{/H]} 
		\leq -0.76~\rm{dex/dex}$, while the range in the major merger 
		simulations is $-0.10\leq \nabla [Z\rm{/H]} \leq -0.46~\rm{dex/dex}$. 
		Within these ranges, both our high-mass bin rGC and bGC metallicity 
		gradients are fully consistent with those from the major merger 
		simulations. 
		However, in the low-mass bin our GC subpopulation metallicity gradients 
		are consistent with both major and minor merger simulations. 
		We will return to these comparisons in more detail in Section 
		\ref{sec:results_fractionaccreted}.

	\subsection{The metallicity offset between inner stars and rGCs}\label{met:offset}
		Thanks to our radially extended stellar and GC spectroscopic metallicity 
		sample we are able to measure the metallicity offset at $1~\rm{R_{e}}$ 
		between stars and rGCs. 

		\placefigE

		We verify the existence of a correlation between the stellar-rGC 
		metallicity difference with galaxy mass from our sample of 12 
		galaxies. 
		In Figure \ref{fig:E} we plot the metallicity difference 
		$\Delta[Z/\rm{H}]$ 	between the rGCs and the stars at $1~\rm{R_{e}}$ for 
		our sample as a function of different galaxy mass proxies. 
		In particular, for the five galaxies with a high number of rGCs (i.e. 
		NGC~1407, NGC~3115, NGC~4278, NGC~4365 and NGC~4649), we have been able 
		to extrapolate the rGC metallicity to $R\approx1~\rm{R_{e}}$ from the 
		best linear fit of the rGC metallicity profile. 
		For such galaxies, the best linear fit parameters are presented 
		in Table \ref{tab:C}.
		Because of the low-number of rGCs in the other galaxies, we adopt the 
		average metallicity of the whole rGC subpopulation in each galaxy as a 
		proxy for the value at $R\approx1~\rm{R_{e}}$.

		\placetabC

		In the first panel we use galaxy central velocity dispersion 
		$\sigma_{\rm{0}}$ from Table \ref{tab:A}. 
		The best fitting linear function is:
		\begin{eqnarray}
			\Delta[Z/\rm{H}] = \left(2.50\pm0.75 \right)\log \sigma_{\rm{0}}- 
				\left(5.71 \pm 1.76  \right).
		\end{eqnarray}
		The rms of this relation is $0.08$ and the Spearman index value (a 
		statistical measure of the correlation between two variables) is  
		$r_{\rm{s}}=0.45$ with a significance of $p=86.2$ per cent.  

		In the second panel we use total stellar mass $M_\star$ from Table 
		\ref{tab:A}, obtaining the best-fit linear relation:
		\begin{eqnarray}
			\Delta[Z/\rm{H}] = \left(0.50\pm0.32 \right)\log M_{\star}- 
				\left(5.42 \pm 3.50  \right).
		\end{eqnarray}
		The rms is $0.24$ and the Spearman index is $r_{\rm{S}}=0.41$ with a 
		significance of $p=80.9$ per cent. 

		Lastly, in the third panel we use total $B$-band magnitude of the host 
		galaxy, as presented in Table \ref{tab:A}. 
		The best-fit relation is:
		\begin{eqnarray}
			\Delta[Z/\rm{H}] = \left(-0.19\pm0.14 \right)\log M_{B}- 
				\left(3.84 \pm 2.81  \right).
		\end{eqnarray}
		% %
		The rms is $0.28$ and the Spearman index is $r_{\rm{S}}=-0.37$ with a 
		significance of $p=77.0$ per cent. 
		% %

		In summary, we find that the difference between the stellar metallicity 
		and the rGC subpopulation metallicity at $1~\rm{R_{e}}$ correlates 
		mildly with the galaxy mass.  
		Thus, the rGC subpopulations in low-mass galaxies show an 
		average metallicity closer to that of the galaxy stars at similar radii 
		than in high-mass galaxies. 

		This trend may suggest a longer lag between the peaks of formation for 
		rGCs and stars in more massive galaxies (i.e. continued enrichment in 
		galaxy stars), as suggested by \citet{Spitler10}. 
		Supporting this view, \citet{Montes14b} found that the rGCs in M87 are 
		older and more metal-poor than the bulk of the stars in the central 
		regions of the galaxy. 
		However, as lower-mass galaxies have been found to experience 
		a more extended star formation than their higher-mass 
		counterparts \citep{Thomas05, McDermid15}, the trend between stellar-rGC 
		metallicity difference and galaxy mass should be the opposite (i.e. 
		larger metallicity offsets in lower mass galaxies). 

		However, there is another possible cause for this correlation. 
		The correction applied to our stellar metallicities is obtained 
		from the offsets with \sauron\ metallicities (see Section 
		\ref{sec:stellarMet}). 
		Since this offset is probably linked with a more bottom-heavy IMF in 
		the central regions of high-mass galaxies \citep{Pastorello14}, this may 
		not apply to the region at $R\ge1~\rm{R_{e}}$, where the IMF has been 
		found to be shallower and similar among galaxies with different mass 
		\citep{MartinNavarro15}. 
		In this case, the stellar metallicity at $1~\rm{R_{e}}$ in high-mass 
		galaxies may be systematically overestimated. 

		\citet{Goudfrooij13} found that a colour offset between rGCs and stars 
		in high-mass galaxies exists out to around $1~\rm{R_{e}}$. 
		From the Lick indices analysis in two of their galaxies, they found 
		consistent ages and metallicities between rGCs and stars. 
		Therefore, they claimed that only a steep IMF for the galaxy stars can 
		explain the colour offset with the rGCs. 
		This is in conflict with our findings at $1~\rm{R_{e}}$, since we 
		spectroscopically measure different metallicities for rGCs and stars. 
		However, as discussed above, a shallower IMF slope in high-mass galaxies 
		at $1~\rm{R_{e}}$ may reduce the stellar metallicity at such radii. 

	\subsection{Summary of the results}

		In summary, we find that the possible metal-poor stellar population 
		at large radii from RGB stellar studies shows a metallicity similar to 
		that of bGCs. 
		In addition, we observe that the metallicity gradients of bGCs and rGCs 
		are consistent one with another within the uncertainties in the same 
		galaxy mass bin. 
		Moreover, in the high-mass galaxy bin, both bGCs and rGCs show 
		shallower metallicity gradients than in the low-mass galaxy bin. 
		We also compare our metallicity profiles in low- and high-mass 
		galaxies with stellar metallicity profiles from simulations assuming 
		a minor merger and a major merger dominated evolutions. 
		We find that the bGC metallicity gradient in our low-mass galaxies is 
		consistent with the former, while in the high-mass galaxies both rGC and 
		bGC metallicity gradients are consistent with the latter. 
		Moreover, we observe that the average stellar metallicity at 
		$1~\rm{R_{e}}$ is very similar to that of the rGC subpopulation, with 
		larger differences for higher-mass galaxies, where the metallicity of 
		the stars is higher than that of the rGCs. 
		In the few cases where we can probe the stellar component to large 
		$\rm{R_{e}}$ (i.e. NGC~3115, NGC~3377, NGC~4473 and NGC~4494), we find 
		that the stellar metallicity at large radii tends to be similar to the 
		average metallicity of the bGC subpopulation. 

\section{Discussion}\label{sec:discussion}

	\subsection{Stellar metallicity profiles at large radii}\label{sec:results_twophase}

		In what follows, we discuss the results within the framework of 
		the two-phase galaxy formation scenario, although the general 
		conclusions we draw from our observations can be used as constraints for 
		future different galaxy formation models. 

		In a two-phase formation scenario, the outer regions of massive ETGs are 
		thought to be dominated by stars formed ex-situ, i.e. accreted from 
		smaller companion galaxies or during major merger events. 
		In the first case, the relation between a galaxy's mass and its mean 
		stellar metallicity dictates that small accreted galaxies are made up of 
		metal-poor stars. 
		Therefore, if the stars in the outskirts of ETGs are mostly accreted 
		from smaller satellite galaxies, these regions should reveal a stellar 
		population with a low mean metallicity. 
		In this case, higher metallicity inner regions and lower metallicity 
		outskirts in the final galaxy are expected to be separated by a 
		transition region. 
		The metallicity profile in this region will be steeper than the original 
		one at the same radii \citep{Hirschmann15}. 
		However, quantitative predictions about the galactocentric radius of 
		this region vary in the literature. 
		One issue is the estimation of the effects of SNe and AGN feedback on 
		the final stellar metallicity profiles, even though some recent studies 
		have started to tackle this problem in detail (e.g. 
		\citealt{Hirschmann15}). 

		In the case of major mergers, accreted stars will have a 
		metallicity comparable with that of pre-existing in-situ formed stars. 
		Thus, any pre-existing metallicity gradient will be washed out 
		by the major merger event \citep{Hirschmann15}. 
		In addition, a significant fraction of in-situ stars is pushed at large 
		radii, further diluting their metallicity gradient. 

		In \citet{Pastorello14} we found that the stellar metallicity outside 
		$1~\rm{R_{e}}$ declines steeply with radius in low-mass galaxies, 
		while it displays a shallower gradient in high-mass systems. 
		This is either a reflection of a relative lack of 
		mergers in the assembly history of our lower-mass ETGs and/or we are 
		probing the region between in-situ and ex-situ dominated regions,
		where the latter are built-up via minor mergers. 
		If it is a lack of mergers, then the galaxy metallicity profiles are 
		still similar to those formed during the dissipative phase of galaxy 
		formation. a
		If, instead, minor mergers dominate in the evolutionary history of 
		low-mass galaxies, then a stellar low-metallicity ``plateau'' should be 
		present at larger radii, where the accreted stars dominate the stellar 
		population.  

		Spectroscopically probing such regions is still a challenge, but 
		some hints of the stellar metallicity profiles at large radii come from 
		resolved RGB stars in nearby low-mass galaxies, albeit with some 
		limitations (e.g. limited radius baseline, observational completeness 
		issues). 
		For example, the almost radially constant RGB stars' metallicities in 
		NGC~3115 \citep{Elson97,Peacock15} suggest that the average integrated 
		stellar metallicity in these outer regions does indeed reach a 
		``plateau'' for $R>1\rm{R_{e}}$. 
		The RGB metallicities obtained for NGC~3377 at large 
		radii by \citet{Harris07a} also show a flat metallicity radial profile. 
		On the other hand, in the nearby galaxy NGC~5128, RGB 
		metallicities have been found to show a mild gradient (i.e. 
		$\Delta[Z/\rm{H}]/\Delta R_{\rm{e}}\approx -0.030\pm0.003~\rm{dex}$) 
		%Rejkuba+14
		out to more than $20~\rm{R_{e}}$ \citep{Rejkuba14,Bird15}. 
		
		An interesting addition to this picture is the observation of 
		distinct stellar components in the data of both \citet{Elson97} and 
		\citet{Peacock15}. 
		Their median metallicities are qualitatively close to those of the red 
		and blue GCs in NGC~3115 at the same radii. 

	\subsection{What is the origin of GC metallicity bimodality?}\label{sec:results_bimodality}
		Most GC systems of large galaxies show a bimodality in colour, which 
		generally reflects a bimodality in metallicity (\citealt{Brodie06, 
		Brodie12, Usher12} and references therein). 
		Most current theoretical models of GC formation predict that rGCs are 
		mostly formed in-situ, together with the bulk of the host galaxy's stars, 
		and bGCs form in dwarf galaxies that are later accreted, preferentially 
		in the outer regions of the larger galaxies \citep{Cote98,Tonini13, 
		Katz13, Katz14}, although some in-situ GC formation in the 
		galaxy halo may occur \citep{Canning14}.
		In this scenario, the bGC average metallicity is lower than that of rGCs 
		because it reflects that of the low-mass systems where the bGCs form. 

		However, another possible scenario for the bimodality formation allows 
		both rGCs and bGCs to be formed in-situ and also accreted from low-mass 
		galaxies. 
		In this case, in-situ rGCs and bGCs mostly populate the inner regions of 
		the host galaxy, while the accreted GCs lie in the outer regions.  
		If the main sources of ex-situ GCs are minor mergers, one may expect a 
		lower number of accreted rGCs than accreted bGCs. 
		This because low-mass accreted galaxies likely had low numbers of 
		rGCs, hence their contribution to the host galaxy's rGC subpopulation 
		should be small \citep{Forbes97,Strader06,Peng06,Forbes11}.  
		On the other hand, major mergers will contribute with a more
		similar number of rGCs and bGCs. 

		The most notable difference between the two scenarios concerns the 
		origin of the bGCs, for which the available theoretical models predict 
		almost exclusively ex-situ formation, while in the second case a 
		significant fraction of bGCs may have formed in-situ.  

		Differences in rGCs and bGCs azimuthal, kinematic and metallicity 
		distributions are expected in the two scenarios. 

		If bGCs are exclusively formed ex-situ and later accreted, their 
		azimuthal and kinematic distributions should be, to first order,
		decoupled from those of the rGCs and the galaxy inner stars. 
		Instead, if bGCs and rGCs formed under similar conditions, both 
		will show in the inner regions kinematics and azimuthal distributions 
		comparable to those of the galaxy stars (i.e. with similar ellipticity 
		and PA). 
		Observations of GC kinematics and azimuthal distributions do not give a
		clear answer to the origin of bGCs. 
		In fact, several 
		observations favour the latter scenario (i.e. bGCs and rGCs formed 
		under similar conditions), with galaxies displaying azimuthal 
		distributions for both rGCs and bGCs similar to those of the host 
		galaxy's stellar component (e.g. \citealt{Dirsch05, 
		Hargis14}), while others have more decoupled components (e.g. 
		\citealt{Strader11, Kartha14}). 
		Moreover, in a number of nearby ETGs the kinematics of bGCs have been 
		found to be significantly distinct from those of the stellar component 
		and the rGCs (e.g. \citealt{Schuberth10, Pota13}). 

 		A better way to disentangle the contribution of in-situ and ex-situ 
		formed GCs in the two subpopulations is to examine the relation 
		between their average metallicity and host galaxy mass. 
		If a GC subpopulation exclusively formed in-situ, then its metallicity 
		should show a positive correlation with host galaxy mass, i.e. in 
		high-mass galaxies the in-situ formed GCs will have higher average 
		metallicities than those in low-mass galaxies. 
		% %
		On the other hand, the average metallicity of an accreted component 
		would show a weaker correlation with the \textit{current} host 
		galaxy mass. 
		This is because the mass of an accreted satellite is not directly 
		dependent on the mass of the final host galaxy, although high mass 
		galaxies tend to accrete higher mass satellites \citep{Oser12}. 
		So, if rGCs are mostly formed in-situ and bGCs are accreted, 
		the rGCs will show a strong trend of their average metallicity with the 
		host galaxy mass, while for the bGCs this trend will be 
		weak/negligible \citep{Bekki08}. 

		It is not clear whether or not the bGC average metallicity trend with 
		host galaxy mass is consistent with this picture. 
		A weak correlation between bGC average colour/metallicity and 
		galaxy mass has been confirmed by several studies 
		(e.g. \citealt{Larsen01, Strader04, Strader06, Peng06}).  
		However, in the cases where a colour/metallicity gradient exists (e.g. 
		\citealt{Peng06}) this result may be simply a consequence of an aperture 
		bias (i.e. a GC system average will lead to measuring the 
		colours/metallicities of mostly the innermost bGCs in massive galaxies). 
		Once corrected by this bias, the estimated slope for the relation 
		between bGC average metallicity and host galaxy mass may be greatly 
		reduced, potentially disappearing, as discussed in \citet{Liu11}. 
		This would reinforce the theoretical view of exclusively externally 
		formed bGCs. 
		In our low-mass galaxy bin we note that bGCs show a lower 
		average metallicity (i.e. $\overline{[Z/\rm{H]}}=-1.21\pm0.04~\rm{dex}$) 
		than in the high-mass galaxy bin (i.e. $\overline{[Z/\rm{H]}}=-1.12\pm
		0.03~\rm{dex}$) if compared over the same radial range (i.e. $1\leq R/
		\rm{R_e}\leq8$). 
		Since our spectroscopic sample is not affected by aperture bias, a 
		mild trend of bGC average metallicity with galaxy mass may exist. 

		Another key difference between in-situ and ex-situ formed 
		GCs concerns their radial colour/metallicity gradients. 
		Similar to the predictions for the stellar component, in-situ formed GCs 
		will be more metal-rich closer to the centre of the gravitational well. 
		This is because the gas from which GCs form is more metal-rich where 
		the stellar evolution is faster. 

		\citet{Hirschmann15} found that the stellar metallicity profile in the 
		region $2\le R \le6~\rm{R_{e}}$ is steeper in the minor merger 
		simulations, since the accreted stars at large radii will have a lower 
		metallicity than the in-situ formed stars at the same radii. 
		In general, one may expect a similar trend for the case of in-situ and 
		ex-situ GCs, with steeper gradients if these latter are accreted during 
		minor merger episodes. 
		However, the specific frequency of rGCs is lower in low-mass galaxies 
		and, therefore, minor merger events will generally contribute with more 
		bGCs than rGCs. 
		Major mergers, instead, will strongly reduce the steepness of the 
		metallicity profiles at almost any radius. 
		This is because the accreted GCs will be radially mixed in the host 
		galaxy and the in-situ formed GCs will be pushed at larger radii.
		However, since the GCs formed in more massive satellites tend to be 
		accreted at smaller galactocentric radii, a mild metallicity gradient 
		may be created by this mass segregation effect.   

		Therefore, if rGCs are mostly formed in-situ and bGCs are formed 
		ex-situ, the former should show steeper metallicity gradients than the 
		latter. 
		Instead, if the two GC subpopulations have had a similar formation 
		process, both being formed in-situ and accreted, they are expected to 
		show similar metallicity gradients over all radii (i.e. steep in the 
		inner regions and shallow in the outer regions). 

		In the literature there are several massive ETGs for which both 
		the rGCs and bGCs in the inner regions show a significant 
		colour/metallicity gradient, while in the outer regions such 
		gradients flatten \citep{Forte01, Harris09b, Harris09c, 
		Forbes11, Arnold11, Liu11, Jennings14, Hargis14, Pota15}. 
		While this strongly supports a common in-situ formation for the two 
		GC subpopulations in the inner regions, the metallicities have been 
		obtained from photometric colours. 
		Even though most GCs have old ages, these colour gradients may 
		include age variations, thus obscuring the true metallicity 
		gradients. 
		In summary, from the available literature it is not clear whether 
		bGCs may be also formed in-situ or are exclusively accreted (as 
		suggested by most theoretical models). 

		In Section \ref{sec:results_ourresults} we presented the spectroscopic 
		metallicity gradients for both rGCs and bGCs in two galaxy mass bins. 
		In both mass bins, the gradients of the two GC subpopulations are 
		similar to each other. 
		These results support a scenario in which rGCs and bGCs in the 
		same galaxy share a common origin. 
		In particular, the significant bGC and rGC metallicity gradients in our 
		low-mass galaxy bin are in agreement with a mostly in-situ origin for 
		both the GC subpopulations.

	\subsection{Fraction of accreted GCs}\label{sec:results_fractionaccreted}
		As seen in the previous Section, from the rGC and bGC metallicity 
		gradients it is possible to infer the existence of both in-situ and 
		ex-situ formation for both GC subpopulations. 
		Under the two-phase galaxy formation scenario, the ex-situ fraction 
		of stars is predicted to be higher in more massive galaxies than in 
		their low-mass counterparts (e.g. \citealt{Oser10, Hirschmann13}).  
		The \citet{Hirschmann15} simulations, including stellar wind feedback, 
		also showed that the ex-situ stars accreted during minor merger episodes 
		contribute to steepening the pre-existing in-situ stellar metallicity 
		gradient. 
		In addition, major mergers create a shallower stellar metallicity 
		profile in the outer regions. 
		If these trends with the formation history apply also to the GC 
		subpopulations, one can draw some simplistic conclusions from the GC 
		metallicity gradients in galaxies of different mass. 
		In particular, if we assume that GCs are a good proxy for the stellar 
		component at large radii, then we can compare the \citet{Hirschmann15} 
		stellar metallicity gradient predictions with our GC metallicity 
		gradients. 
		A caveat to keep in mind is that \citet{Hirschmann15} simulations regard 
		galaxies with $M>10^{11}~\rm{M_{\odot}}$, and thus our low-mass galaxies 
		are not directly comparable with their simulations. 
		However, in what follows we assume that \citet{Hirschmann15} predictions 
		are also valid for the galaxies in our low-mass bin (see Section 
		\ref{comparisonHirschmann}). 

		Similarly to the stellar component, if in high-mass galaxies a high 
		fraction of rGCs and bGCs are accreted during major mergers, then the 
		final rGC and bGC metallicity profiles should be flat. 
		Conversely, if most of the GCs form in-situ or are accreted during minor 
		mergers, then they should show steep metallicity gradients. 
		In particular, if the ex-situ GCs are accreted during minor mergers 
		episodes, then the bGC metallicity profiles may be expected to be 
		steeper than those of the rGCs in the same galaxies. 
		This is because, given the different specific frequency of rGCs and 
		bGCs in low-mass systems, minor merger episodes will in general 
		contribute with more accreted bGCs than rGCs. 
		As a consequence, the total metallicity gradients of both GC 
		subpopulations will be steeper than in the major merger accretion case, 
		but rGCs will also show a shallower gradient than the bGCs. 

		This is consistent with our results from Section 
		\ref{sec:results_ourresults}. 
		The shallower (consistent with being flat) metallicity gradients we  
		measure for the rGCs and bGCs in high-mass galaxies are consistent 
		with the presence of a higher fraction of ex-situ formed GCs accreted 
		during major mergers in these systems.  
		We find that both these metallicity gradients are consistent with the 
		stellar metallicity gradients obtained in a major merger dominated 
		evolutionary history in the simulation by \citet{Hirschmann15}. 

		On the other hand, the steeper metallicity gradients for rGCs 
		and bGCs in low-mass galaxies show consistency with both  
		minor and major merger dominated formation histories. 
		This is qualitatively similar to that observed for the stellar component 
		in low-mass galaxies \citep{Pastorello14}.
		The rGC metallicity gradient in the same galaxy mass bin is shallower. 
		We argue that this is because the number of accreted rGCs is lower than 
		for the bGCs and, therefore, their metallicity gradient is more similar 
		to the in-situ stellar metallicity gradient.  

		In Figure \ref{fig:F} we present a further comparison of our GC 
		metallicity gradients with the simulations by \citet{Hirschmann15} 
		that include stellar wind feedback. 
		In particular, we plot the simulated gradients for both of the minor 
		and major merger models against the mass weighted merger ratio and 
		the fraction of ex-situ stars in the simulations. 

		In general, the simulated minor merger stellar metallicity gradients 
		have lower mass weighted merger ratios and lower fractions of ex-situ 
		stars than the simulated major merger models. 
		In the same Figure, we also show the average metallicity gradients of 
		rGCs and bGCs in our two mass bins. 
		In the low-mass galaxy bin, the average metallicity for the two GC
		subpopulations is qualitatively in between those of the minor 
		and major merger models.  
%		although with a large uncertainty. 
		%
		We note that, even though the \citet{Hirschmann15} simulations 
		did not include galaxies in our low mass bin mass range, the effects of 
		the formation history on the metallicity gradients is the dominant 
		factor. 
		In the high mass galaxy bin, the average metallicity gradients 
		of both the two GC subpopulations are similar to those from 
		the major merger models.

		\placefigF

%%%%%%%%%%%%%%%%%%%%%%%
%												%
%  6. Conclusions						%
%  \label{sec:conclusions}		%
%												%
%%%%%%%%%%%%%%%%%%%%%%%

\section{Conclusions}
\label{sec:conclusions}
	Here we present for the first time a systematic compilation of 
	spectroscopically measured metallicities for the stellar and the globular 
	cluster components of 12 galaxies as part of the SLUGGS survey. 
	Metallicities at large radii for both stars and globular clusters have been 
	obtained from the CaT index. 
	For the innermost radii, we adopt stellar metallicities from \sauron\ 
	\citep{Bacon01} and \atlas\ \citep{Cappellari11a} surveys for the galaxies 
	in common. 
	Combining \sauron, \atlas\ and SLUGGS values, we have been able to 
	spectroscopically measure the metallicity of the stellar component out to 
	almost $4~\rm{R_{e}}$, and of the red and blue globular clusters out to 
	almost $15~\rm{R_{e}}$. 

	Our main results can be summarized as follows:

	\begin{itemize}

		\item{	We spectroscopically confirmed that some red globular clusters 
				exist out to $R>10~\rm{R_{e}}$. 
				}

		\item{	We compare the average stellar metallicity at $1~\rm{R_{e}}$ 
				with that of the red globular cluster component, finding that 
				high-mass galaxies host a stellar population that is more 
				metal-enriched than the red globular cluster component, while 
				in low-mass galaxies the stellar and the red globular cluster 
				metallicities are more similar. 
				We claim that this can not be simply due to a bottom-heavy IMF. 
				}

		\item{	We find that the metallicity gradients of red and blue globular 
				clusters are consistent within the same galaxy mass bin.  
				In particular, in the low galaxy mass bin blue and red globular 
				cluster metallicity gradients are $-0.26\pm0.08$ and $-0.45\pm
				0.13~\rm{dex/dex}$, respectively. 
				In the high galaxy mass bin, red and blue globular cluster 
				metallicity gradients are $-0.07\pm0.08$ and $-0.08\pm0.09~
				\rm{dex/dex}$, respectively. 
				This consistency between globular cluster subpopulation 
				metallicity gradients within the same galaxy mass bin suggests a 
				common formation process for both red and blue globular clusters 
				in the same systems. 
				The colour/metallicity bimodality is then potentially a 
				consequence of different epochs for the two subpopulations, as 
				already concluded by photometric studies (e.g. 
				\citealt{Arnold11, Forbes11}). 
				This is in contrast with most of the current theoretical models 
				that consider red globular clusters as mostly formed in-situ and 
				blue globular clusters as mostly formed ex-situ and later 
				accreted into their present day host galaxy. }

		\item{	We measure metallicity gradients for both the GC subpopulations 
				that are steeper in the low galaxy mass bin than in the high 
				galaxy mass bin. 
				From the comparison with stellar metallicity gradients from 
				simulations at similar radii, we find that this trend is 
				qualitatively consistent with a mass-dependent galaxy formation 
				history, in which the ex-situ fraction and the importance of 
				major mergers increase with the galaxy mass. 
				In particular, in high-mass galaxies the higher number of 
				accreted globular clusters yields shallower metallicity 
				gradients because of the dynamical mixing of in-situ and ex-situ 
				formed globular clusters. 
				On the other hand, the blue globular cluster metallicity 
				gradient is consistent with the steep stellar metallicity 
				gradient from simulations of minor merger dominated galaxy 
				formation, while the red globular cluster metallicity gradient 
				in the same galaxies is qualitatively shallower. 
				This is in agreement with a galaxy formation dominated by minor 
				mergers, since the merger of lower-mass satellites will accrete 
				more blue than red globular clusters. 
			}

		\end{itemize}

\placetabAppData

\section*{Acknowledgements}
	We want to thank the (anonymous) referee for useful comments and suggestions 
	which helped to improve this paper. 
	We wish also to thank Luca Cortese, Martina Fagioli, Alister Graham, William 
	Harris, Michaela Hirschmann and Chiara Tonini for the helpful discussions 
	and suggestions. 
	We are also grateful to Harald Kuntschner for sharing NGC~4365 \atlas\ 
	metallicity profile and Mark Peacock for sharing the NGC~3115 RGB stellar 
	metallicity distribution data. 
	Some of the data presented herein were obtained at the W. M.
	Keck Observatory, operated as a scientific partnership among the
	California Institute of Technology, the University of California and
	the National Aeronautics and Space Administration, and made possible 
	by the generous financial support of the W. M. Keck Foundation. 
	The authors wish to recognize and acknowledge the very significant 
	cultural role and reverence that the summit of Mauna Kea
	has always had within the indigenous Hawaiian community. 
	The analysis pipeline used to reduce the DEIMOS data was developed
	at UC Berkeley with support from NSF grant AST-0071048. 
	DF thanks the ARC for support via DP130100388. 
	This work was supported by NSF grant AST-1211995.

\bibliographystyle{mn2e}
\bibliography{bibliography}{}

%%%%%%%%%%%%%%%%%%%%%%%
%												%
%  A. Appendix						%
%  \label{sec:appendix}		%
%												%
%%%%%%%%%%%%%%%%%%%%%%%

\appendix
\label{sec:appendixA}	
\section{Updated stellar metallicity maps and radial profiles}
	Since the publication of \citet{Pastorello14}, we have obtained new stellar 
	metallicity data from the ongoing SLUGGS survey. 
	In particular, for the purposes of this work we have updated the metallicity 
	dataset for NGC~3115. 
	In Figure \ref{fig:AppA2} we present the new metallicity profiles and 2D 
	map. 
	An important difference between this and the \citet{Pastorello14} 
	metallicity maps for NGC~3115 is that the higher metallicity substructure 
	previously extracted in the North-East region region of the galaxy is not 
	visible anymore. 
	The relatively low number of datapoints in the old version of the 2D 
	kriging map biased the results, in the sense that few high metallicity 
	datapoints in the same region caused an overestimation of the kriging 
	extracted metallicity. 

	\placefigAppendixGalaxy

\end{document}